\documentclass[preprintnumbers,prd,twocolumn,tightenlines,floatfix,showpacs,amssymb,nofootinbib,superscriptaddress]{revtex4}

\usepackage{pictex}
\usepackage[dvips]{graphicx}
\usepackage{amsmath}

\usepackage{amsmath}
\usepackage{amsfonts}
\usepackage{amsthm}

\newcommand{\del}{\partial}
\DeclareMathOperator{\Tr}{Tr}
\DeclareMathOperator{\re}{Re}
\newcommand{\Dslash}{\!\!\!\!/\,}
\newcommand{\pslash}{\!\!\!/}
\newcommand{\nn}{\nonumber}
\newcommand{\half}{\frac{1}{2}}
\newcommand{\quarter}{\frac{1}{4}}
\newcommand{\eqn}[1]{ \begin{equation} #1 \end{equation} }

\newcommand{\link}{
\setlength{\unitlength}{14pt}
\begin{picture}(1,1)(0,0)
\linethickness{0.25pt}
\put(0,0){\circle*{0.15}}
\put(0,0){\vector(1,0){1}}
\put(1,0){\circle{0.14}}
\end{picture}}
\newcommand{\stapleup}{
\setlength{\unitlength}{14pt}
\begin{picture}(1,1)(0,0)
\linethickness{0.25pt}
\put(0,0){\circle*{0.15}}
\put(0,0){\vector(0,1){1}}
\put(0,1){\vector(1,0){1}}
\put(1,1){\vector(0,-1){1}}
\put(1,0){\circle{0.14}}
\end{picture}}
\newcommand{\stapledown}{
\raisebox{-14pt}{
\setlength{\unitlength}{14pt}
\begin{picture}(1,1)(0,-1)
\linethickness{0.25pt}
\put(0,0){\circle*{0.15}}
\put(0,0){\vector(0,-1){1}}
\put(0,-1){\vector(1,0){1}}
\put(1,-1){\vector(0,1){1}}
\put(1,0){\circle{0.14}}
\end{picture}}}

\begin{document}
\preprint{ADP-04-06/T587}

\title{The FLIC Overlap Quark Propagator}

\author{Waseem Kamleh}
\affiliation{Special Research Centre for the Subatomic Structure of Matter and Department of Physics, University of Adelaide 5005, Australia. }
\author{Patrick O. Bowman}
\affiliation{Nuclear Theory Center, Indiana University, Bloomington IN 47408, USA.}
\author{Derek B. Leinweber}
\author{Anthony G. Williams}
\author{Jianbo Zhang}
\affiliation{Special Research Centre for the Subatomic Structure of Matter and Department of Physics, University of Adelaide 5005, Australia. }

\begin{abstract}
FLIC overlap fermions are a variant of the standard (Wilson) overlap action, with the FLIC (Fat Link Irrelevant Clover) action as the overlap kernel rather than the Wilson action. The structure of the FLIC overlap fermion propagator in momentum space is studied, and a comparison against previous studies of the Wilson overlap propagator in quenched QCD is performed. To explore the scaling properties of the propagator for the two actions, numerical calculations are performed in Landau Gauge across three lattices with different lattice spacing $a$ and similar physical volumes. We find that at light quark masses the actions agree in both the infrared and the ultraviolet, but at heavier masses some disagreement in the ultraviolet appears. This is attributed to the two actions having different discretisation errors with the FLIC overlap providing superior performance in this regime. Both actions scale reasonably, but some scaling violations are observed. 
\end{abstract}

\pacs{
12.38.Gc,  
11.15.Ha,  
12.38.Aw,  
14.65.-q   
}

\maketitle

\section{Introduction}

The quark propagator in momentum space is a fundamental quantity in
QCD. Its infrared structure provides insight into the dynamical
generation of mass due to the spontaneous breaking of chiral symmetry
within QCD. Its ultraviolet behaviour allows one to obtain the running
quark mass~\cite{aoki99,becirevic,Bowman:2004xi} and OPE 
condensates~\cite{Arriola:2004en}.  Lattice QCD allows a 
direct probe
of the non-perturbative quark propagator. The momentum-space quark
propagator has been studied previously using different gauge fixing
and fermion actions
~\cite{Bowman:2004xi,jon1,jon2,bowman01,blum01,overlgp,overlgp2},
including studies of the (standard) Wilson overlap in Landau
gauge~\cite{Bowman:2004xi,overlgp,overlgp2}. 
The FLIC overlap~\cite{kamleh-overlap} has been shown to possess computational advantages over the Wilson overlap, and we are interested in comparing the (quenched) Wilson overlap and FLIC overlap propagators to see if there is any difference in their scaling properties.

In the continuum, the tree-level $(A_\mu(x) = 0)$ quark propagator is identified with the (Euclidean space) fermionic Greens function,
\eqn{ (\del\pslash + m^0)\Delta_{\rm f}(x,y) = \delta^4(x-y),}
where $m^0$ is the bare quark mass. In momentum space this equation is solved straightforwardly,
\eqn{ \tilde{\Delta}_{\rm f}(p) = \frac{1}{ip\pslash + m^0}. }
Denote $S^{(0)}(p) \equiv \tilde{\Delta}_{\rm f}(p)$ to be the tree-level propagator in momentum space. Then in the presence of gauge field interactions, define $S_{\rm bare}(p)$ to be the (fourier transform) of the (interacting) fermionic Green's function,
\eqn{ (D\Dslash + m^0)\Delta^{\rm bare}_{\rm f}(x,y) = \delta^4(x-y). }
We define the mass function $M(p)$ and the bare renormalisation function $Z(p)$ such that the bare quark propagator has the form
\eqn{ S_{\rm bare}(p) = \frac{Z(p)}{ip\pslash + M(p)}. }
Then for the renormalisation point $\zeta,$ the renormalised quark propagator is given by
\eqn{ S_\zeta(p) = \frac{Z_\zeta(p)}{ip\pslash + M(p)} = Z_2(\zeta,a)S_{\rm bare}(p), }
where $Z_\zeta(p)$ is the ($\zeta$-dependent) renormalisation function, and $Z_2(\zeta,a)$ is the wave function renormalisation constant, which depends on $\zeta$ and the regulator parameter $a.$ The $a$-dependence of $S_{\rm bare}$ is implicit. $Z_2(\zeta,a)$ is chosen such that
\eqn{  Z_\zeta(\zeta) = 1. }
As $S_\zeta(p)$ is multiplicatively renormalisable, all of the $\zeta$-dependence is contained within $Z_\zeta(p),$ that is, the mass function is $\zeta$-independent.

\section{Overlap Fermions}

The overlap formalism~\cite{overlap1,overlap2,overlap3,overlap4} in the vector-like case leads to the following definition of the massless overlap-Dirac operator~\cite{neuberger-massless},
\eqn{ D_{\rm o} = \frac{1}{2a}(1 + \gamma_5 \epsilon(H)).}
Here, $\epsilon(H)$ is the matrix sign function applied to the overlap kernel $H.$ The kernel can be any Hermitian version of the Dirac operator which represents a single fermion species of large negative mass. The standard choice is the Hermitian Wilson-Dirac operator (setting $a=1$),
\eqn{ H_{\rm w} = \gamma_5(\nabla\Dslash + \frac{1}{2}\Delta - m_{\rm w}),}
where $\nabla\Dslash$ is the central covariant finite difference operator, and $\Delta$ is the lattice Laplacian, or Wilson term. In terms of the parallel forward (backward) transport operators 
\begin{align}
T_\mu\psi(x) &= U_\mu(x)\psi(x+e_\mu), \\
T_\mu^\dagger\psi(x) &= U_\mu^\dagger(x-e_\mu)\psi(x-e_\mu),
\end{align}
we have
\begin{align}
\nabla\Dslash &= \half \sum_\mu \gamma_\mu(T_\mu - T_\mu^\dagger), \\
\Delta &= \sum_\mu 2 - T_\mu - T_\mu^\dagger.
\end{align}
 It has been shown that alternative kernel choices, in particular the FLIC (Fat Link Irrelevant Clover) action~\cite{zanotti-hadron}, can accelerate the evaluation of the overlap-Dirac operator~\cite{kamleh-overlap,kamleh-spin}. The FLIC fermion action is a variant of the clover action where the irrelevant operators are constructed using APE-smeared links~\cite{ape-one,ape-two,derek-smooth,ape-MIT}, and mean field improvement~\cite{lepage-mfi} is performed. The Hermitian FLIC operator is given by
\eqn{H_{\rm flic} = \gamma_5(\nabla\Dslash_{\rm mfi} + \frac{1}{2}(\Delta^{\rm fl}_{\rm mfi} - \frac{1}{2}\sigma\cdot F^{\rm fl}_{\rm mfi}) - m_{\rm w}),}
where the presence of fat (smeared) links and/or mean field improvement has been indicated by the super- and subscripts. In this work we choose $\sigma_{\mu\nu}=\half[\gamma_\mu,\gamma_\nu]$ and use a standard one-loop $F_{\mu\nu},$
\begin{align}
F_{\mu\nu}(x) &= \half(C_{\mu\nu}(x) - C^\dagger_{\mu\nu}(x)), \\
\nn C_{\mu\nu}(x) &= \quarter(U_{\mu,\nu}(x) + U_{-\nu,\mu}(x) \\
&\qquad + U_{\nu,-\mu}(x) + U_{-\mu,-\nu}(x)),
\end{align}
where $U_{\mu,\nu}(x)$ is the elementary plaquette in the $+\mu,+\nu$ direction. 

The APE smeared links $U^{\rm fl}_\mu(x)$ constructed from $U_\mu(x)$ by performing $n$ smearing sweeps, where in each sweep we first perform an APE blocking step,
\begin{equation}
V^{(j)}_\mu(x) = (1-\alpha)\ \link + \frac{\alpha}{6} \sum_{\nu \ne \mu}\ \stapleup\ + \stapledown\ , 
\end{equation}
followed by a projection back into $SU(3), U^{(j)}_\mu(x) = {\mathcal P}(V^{(j)}_\mu(x)).$ In this work, the projection is performed by choosing the matrix which maximises the following gauge invariant measure,
\begin{equation}
U^{(j)}_\mu(x) = \max_{U'\in SU(3)} \re \Tr (U_\mu'(x)V^{(j)\dagger}_\mu(x)).
\end{equation}
As it is only the product $n\alpha$ that matters~\cite{derek-apesmearing}, we fix $\alpha=0.7$ and only vary $n.$

Mean field improvement is performed by making the replacements
\eqn{ U_\mu(x) \to \frac{U_\mu(x)}{u_0},\quad U^{\rm fl}_\mu(x) \to \frac{U^{\rm fl}_\mu(x)}{u^{\rm fl}_0}, }
where $u_0$ and $u_0^{\rm fl}$ are the mean links for the standard and smeared gauge fields. We calculate the mean link via the fourth root of the average plaquette, 
\eqn{ u_0 = \langle {\rm \frac{1}{3}ReTr } U_{\mu\nu}(x) \rangle_{x,\mu<\nu}^{\frac{1}{4}}. }

\section{The Overlap Propagator}

It is easily seen that the continuum massless quark propagator anti-commutes with $\gamma_5,$
\eqn{ \{ \gamma_5, S^{\rm c}_{\rm bare}(p)\big|_{m^0=0} \} = 0. }
A straightforward consequence of the Ginsparg-Wilson relation is the inverse of the overlap Dirac operator satisfies
\eqn{ \{ \gamma_5, D_{\rm o}^{-1} \} = 2\gamma_5. }
Using either a Wilson or FLIC kernel, and noting that the mean link is a function of $a$ and that $u_0(a), u^{\rm fl}_0(a) \to 1$ as $a \to 0,$ we obtain
\eqn{\lim_{a\to 0} D_{\rm o} = \frac{1}{2 m_{\rm w}} D\Dslash.}
It is then natural to define the (external) massless bare overlap propagator on the lattice as~\cite{overlap4,edwards-study}
\eqn{ S_{\rm bare}(p)|_{m^0=0} \equiv \frac{1}{2 m_{\rm w}} (D_{\rm o}^{-1} - 1), }
as we then have that 
\eqn{ \{ \gamma_5, S_{\rm bare}(p)\big|_{m^0=0} \} = 0, }
as in the continuum case. 
The massive overlap Dirac operator is given by~\cite{neuberger-almostmassless}
\eqn{ D_{\rm o}(\mu) = (1-\mu)D_{\rm o} + \mu,}
with $|\mu| < 1$ representing fermions of mass $\propto \frac{\mu}{1 - \mu}.$ The massive (external) bare overlap propagator is defined as~\cite{edwards-study}
\eqn{ S_{\rm bare}(p) \equiv \frac{1}{2 m_{\rm w}(1-\mu)}(D_{\rm o}^{-1}(\mu)-1), }
and with the identification 
\eqn{\label{eq:baremass}\mu = \frac{m^0}{2 m_{\rm w}} }
satisfies
\eqn{ S^{-1}_{\rm bare}(p) = S^{-1}_{\rm bare}(p)\big|_{m^0=0} + m^0. }

In order to construct $M(p)$ and $Z(p)$ on the lattice, we first define ${\cal B}(p),{\cal C}_\mu(p)$ such that
\eqn{ S_{\rm bare}(p) = -i{\cal C}\Dslash(p) + {\cal B}(p).}
Then 
\begin{align}
{\cal{C}}_{\mu}(p)&= \frac{i}{n_{\rm s}n_{\rm c}}\Tr[\gamma_\mu{S_{\rm bare}(p)}], \\
{\cal{B}}(p)&= \frac{1}{n_{\rm s}n_{\rm c}}\Tr[S^{\rm bare}(p)],
\end{align}
where the trace is over colour and spinor indices only, and $n_{\rm s}, n_{\rm c}$ specify the dimension of the spinor and colour vector spaces. Now, define the functions $B(p),C_\mu(p)$ such that the inverse of the bare (lattice) quark propagator has the form
\eqn{ S_{\rm bare}^{-1}(p) = iC\Dslash(p) + B(p). }
Then it is easily seen that
\begin{align}
C_\mu &= \frac{{\cal C}_\mu}{{\cal C}^2 + {\cal B}^2}, & B &= \frac{{\cal B}}{{\cal C}^2 + {\cal B}^2},
\end{align}
where ${\cal C}^2 = {\cal C}\cdot{\cal C}.$ 

The kinematical lattice momentum $q_\mu$ is defined such that at tree-level 
\eqn{ (S^{(0)})^{-1}(p) = i q\pslash + m^0,}
that is $q_\mu(p) = C^{(0)}_\mu(p), m^0 = B^{(0)}(p).$ We note that the simple form of these relations is one of the advantages of the overlap, as the absence of additive mass renormalisation prevents the need for having to perform any tree-level correction~\cite{overlgp} (necessary most other fermions actions~\cite{jon1,jon2,bowman01}), outside of identifying the correct momentum variable $q.$ Now, we define $A(p)$ such that
\eqn{ S_{\rm bare}^{-1}(p) = iq\pslash A(p) + B(p).}
The mass function $M(p)$ and renormalisation function $Z(p)$ may then be straightforwardly constructed,
\begin{align}
Z(p) &= \frac{1}{A(p)},& M(p) &= \frac{B(p)}{A(p)}.
\end{align}

\section{Simulation Details}

While the FLIC overlap and Wilson overlap actions are both free of $O(a)$ errors (at both zero and finite quark mass), the actions may in general differ at $O(a^2)$, and as such there may be some difference in their properties at finite $a$. We conduct a comparison between the two actions by calculating the FLIC overlap propagator on the same set of quenched (Luscher-Weisz) lattices as in the Wilson overlap studies~\cite{overlgp,overlgp2}. All the lattices have approximately the same volume, and the details of all three lattices are given in Table \ref{tab:proplattices}.

\begin{table}[t]
\begin{ruledtabular}
\begin{tabular}{ccccccc}
Volume &$\beta$ & $a$ (fm) & $u_{0}$ & $n_{\rm ape}$ & $n_{\rm low}$ & Phys. Vol. (fm$^4$)\\
\hline
$8^3\times{16}$  & 4.286 & 0.200  & 0.8721 & 7 & 20 & $1.6^3\times{3.2}$ \\
$12^3\times{24}$ & 4.60  & 0.120  & 0.8888 & 4 & 15 & $1.44^3\times{2.88}$ \\
$16^3\times{32}$ & 4.80  & 0.096  & 0.8966 & 4 & 5 & $1.536^3\times{3.072}$
\end{tabular}
\caption{Parameters for the three different lattices. All use a tadpole improved Luscher Weisz gluon action. Lattice spacings are set via a string tension analysis incorporating the lattice Coulomb term. Shown is the lattice volume, gauge coupling $\beta,$ lattice spacing, the mean link, the number of APE sweeps, projected modes and the physical volume. All lattices have $m_{\rm w}=1.4.$}
\label{tab:proplattices}
\end{ruledtabular}
\end{table}

Landau gauge is chosen for the gauge fixing. An improved gauge fixing scheme~\cite{bowman2} is used, and a Conjugate Gradient Fourier Acceleration~\cite{cm} algorithm is chosen to perform the gauge fixing. We use periodic boundary conditions in the spatial directions and anti-periodic in the time direction. The matrix sign function is evaluated using the Zolotarev rational polynomial approximation~\cite{chiu-zolotarev}, of degree (typically 8) chosen to give an accuracy of $2.0\times10^{-8}$ within the spectral range of the kernel, after projecting out low-lying modes.

At tree level the FLIC overlap and Wilson overlap propagators have identical behaviour (although the choice of $m_{\rm w}=1.4$ for the FLIC overlap was slightly different from that chosen for the Wilson overlap). The tree level propagator is calculated directly by setting the links and the mean link to unity. The kinematical lattice momentum $q$ is obtained numerically from the tree level propagator, although it could equally well have been obtained from the analytic form for $q$ derived in Ref.~\cite{overlgp}.

Each lattice ensemble consists of 50 configurations. The lattice Green's function is obtained by inverting the FLIC overlap Dirac operator on each configuration using a multi-mass CG inverter~\cite{many-masses}. The Fourier transform is taken to convert to momentum space, and the the bare quark propagator is obtained from the ensemble average. The quark propagator is calculated for 15 different masses. The details of the FLIC overlap parameters used are presented in Table \ref{tab:flicoverprop}.

\begin{table}[h]
\begin{ruledtabular}
\begin{tabular}{ccccc}
 & $\beta=4.286$ & $\beta=4.60$ & $\beta=4.80$ & $m^0$ \\
\hline
$\mu_1$ & 0.00622 & 0.00400 & 0.00305 & 18 \\
$\mu_2$ & 0.01244 & 0.00800 & 0.00610 & 36 \\
$\mu_3$ & 0.01866 & 0.01200 & 0.00915 & 54 \\
$\mu_4$ & 0.02488 & 0.01600 & 0.01220 & 72 \\
$\mu_5$ & 0.03110 & 0.02000 & 0.01525 & 90 \\
$\mu_6$ & 0.03732 & 0.02400 & 0.01830 & 108 \\
$\mu_7$ & 0.04354 & 0.02800 & 0.02134 & 126 \\
$\mu_8$ & 0.04976 & 0.03200 & 0.02439 & 144 \\
$\mu_9$ & 0.06220 & 0.04000 & 0.03049 & 181 \\
$\mu_{10}$ & 0.07464 & 0.04800 & 0.03659 & 217 \\
$\mu_{11}$ & 0.09330 & 0.06000 & 0.04574 & 271 \\
$\mu_{12}$ & 0.12439 & 0.08000 & 0.06098 & 362 \\
$\mu_{13}$ & 0.15549 & 0.10000 & 0.07623 & 452 \\
$\mu_{14}$ & 0.18659 & 0.12000 & 0.09148 & 543 \\
$\mu_{15}$ & 0.21769 & 0.14000 & 0.10672 & 633 
\end{tabular}
\caption{For each $\beta,$ displayed is the 15 lattice masses $\mu.$ The final column shows the bare quark mass in MeV, which is approximately the same for each lattice (see Eq. (\ref{eq:baremass})).}
\label{tab:flicoverprop}
\end{ruledtabular}
\end{table}

\section{Results}

\begin{figure*}[p]
\includegraphics[angle=90,height=0.28\textheight,width=0.45\textwidth]{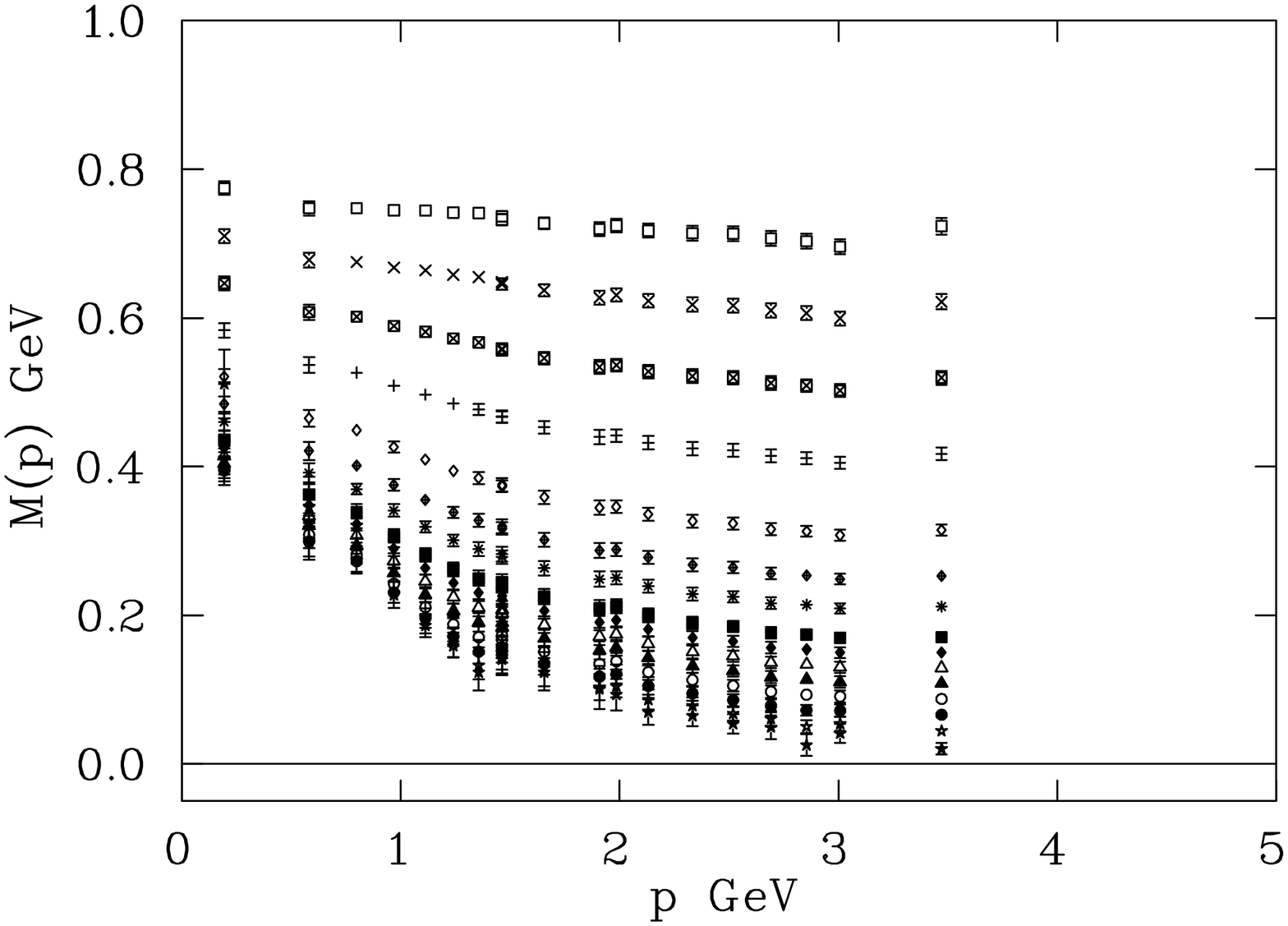}
\includegraphics[angle=90,height=0.28\textheight,width=0.45\textwidth]{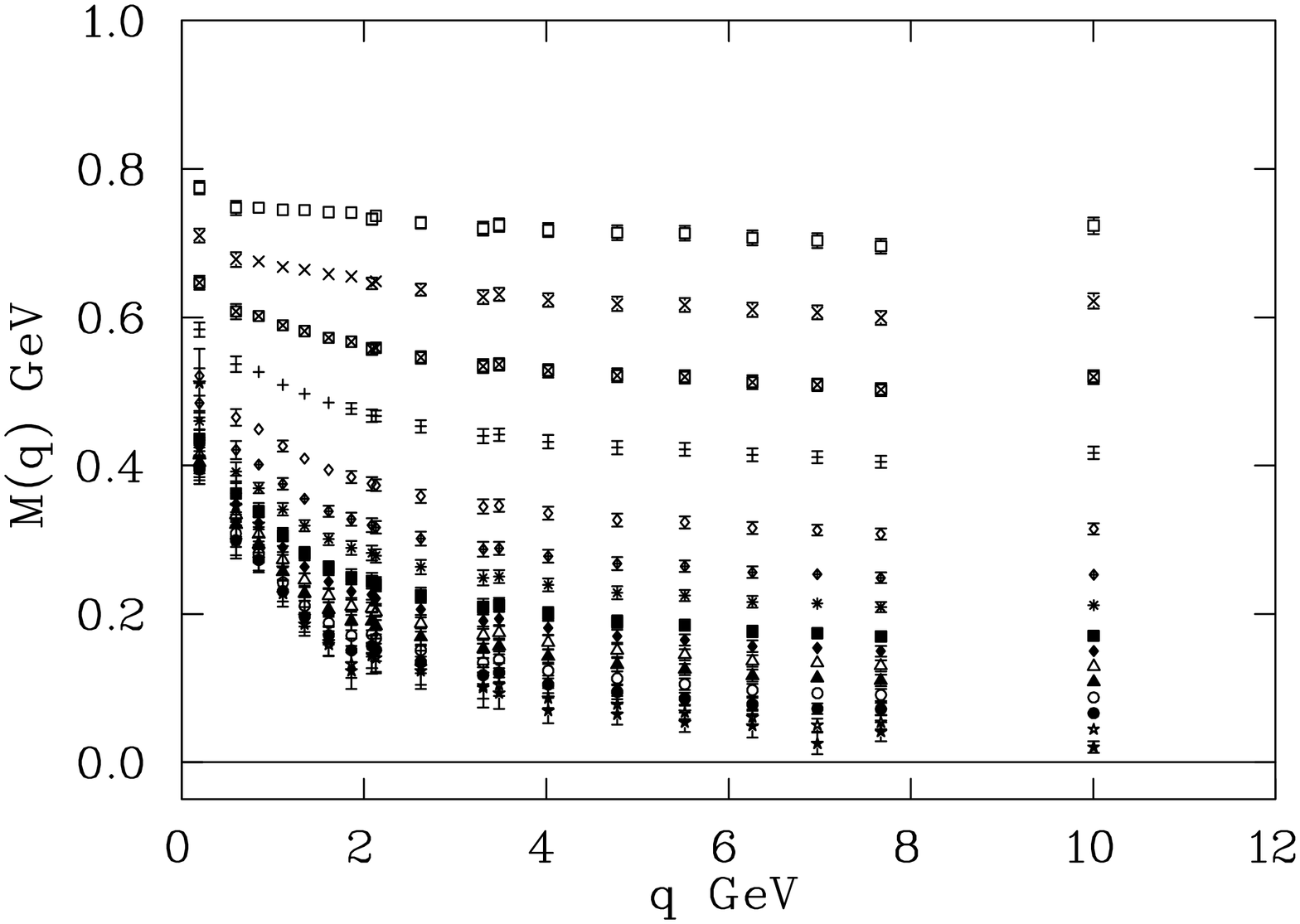}

\includegraphics[angle=90,height=0.28\textheight,width=0.45\textwidth]{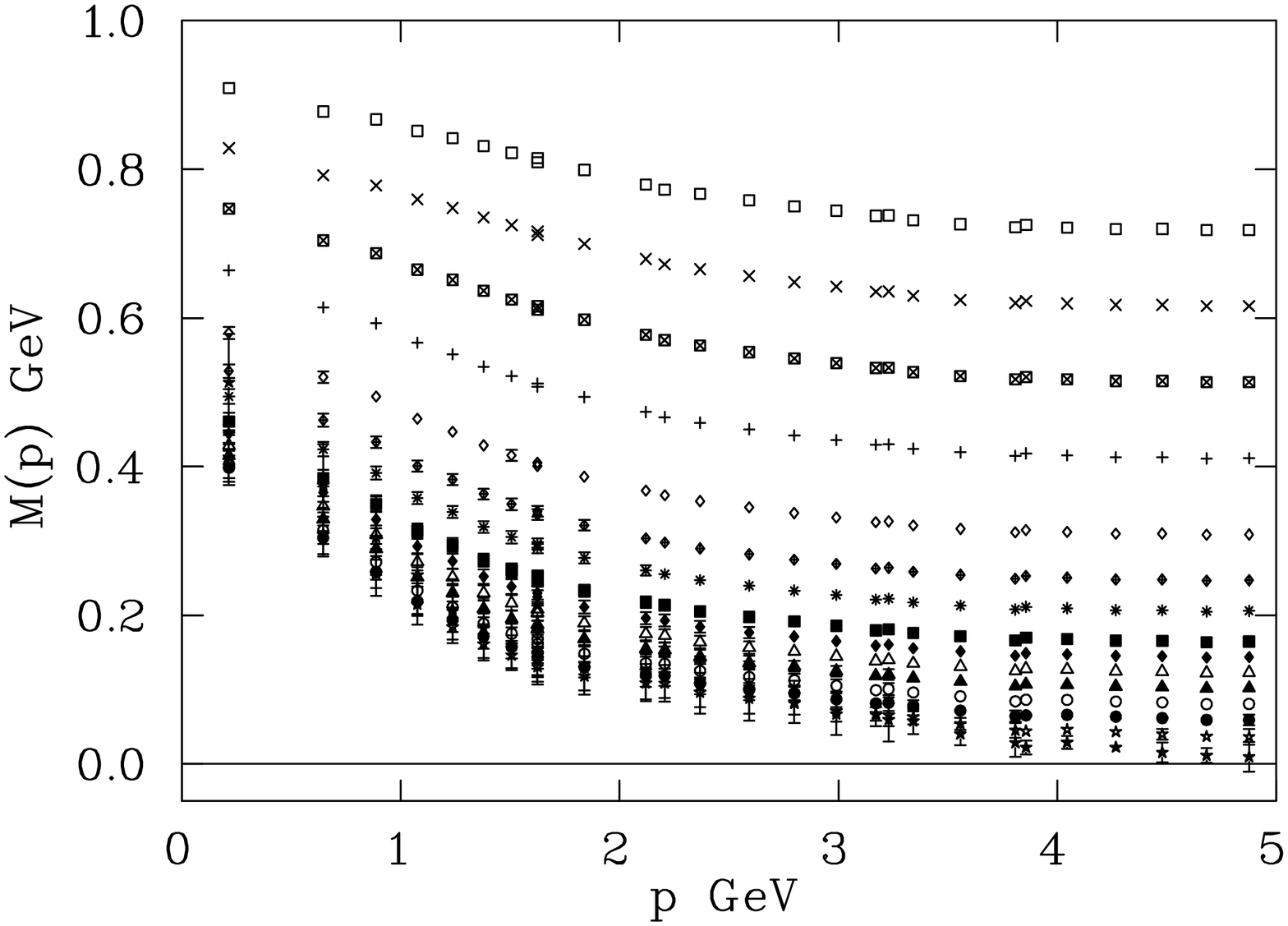}
\includegraphics[angle=90,height=0.28\textheight,width=0.45\textwidth]{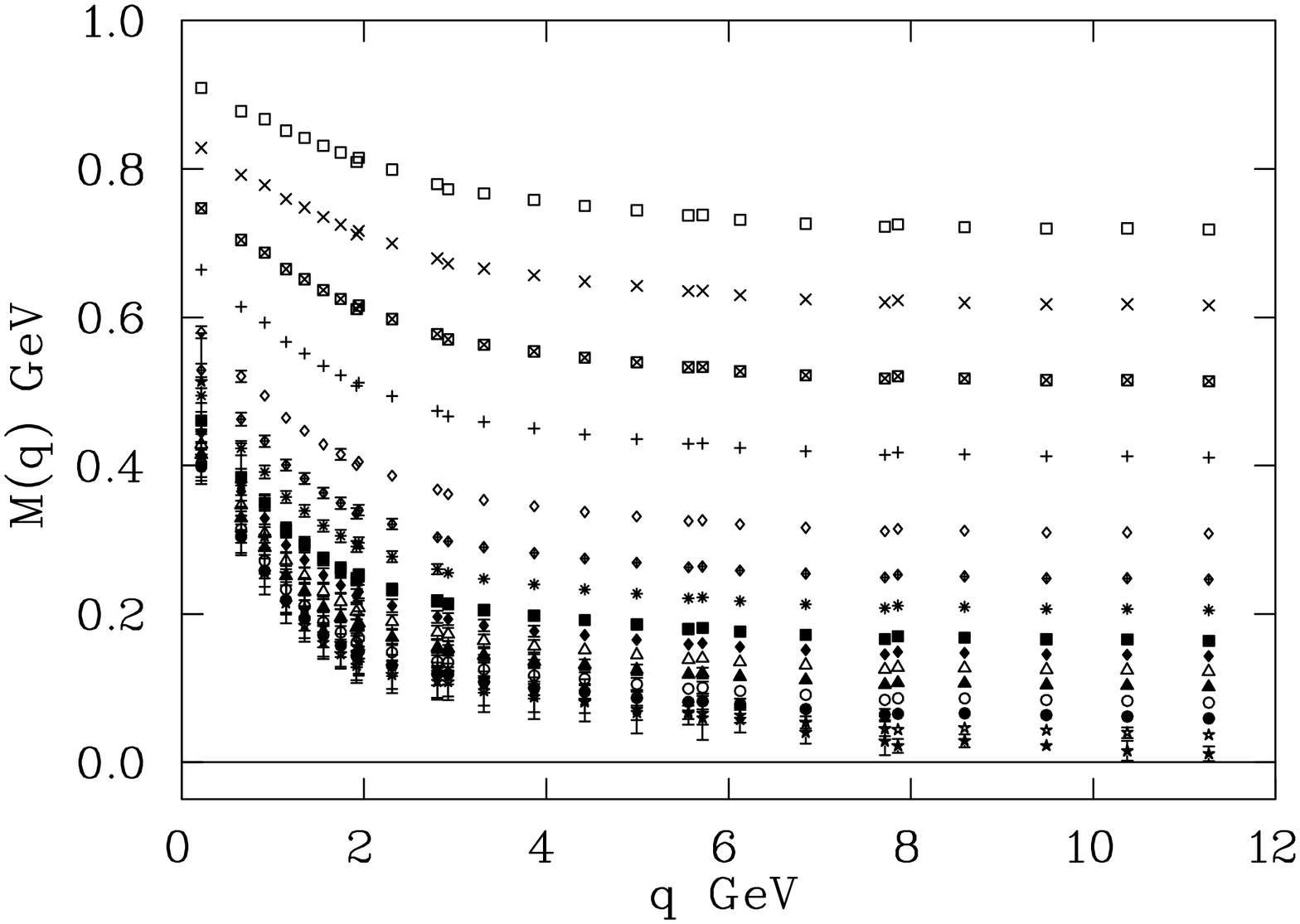}

\includegraphics[angle=90,height=0.28\textheight,width=0.45\textwidth]{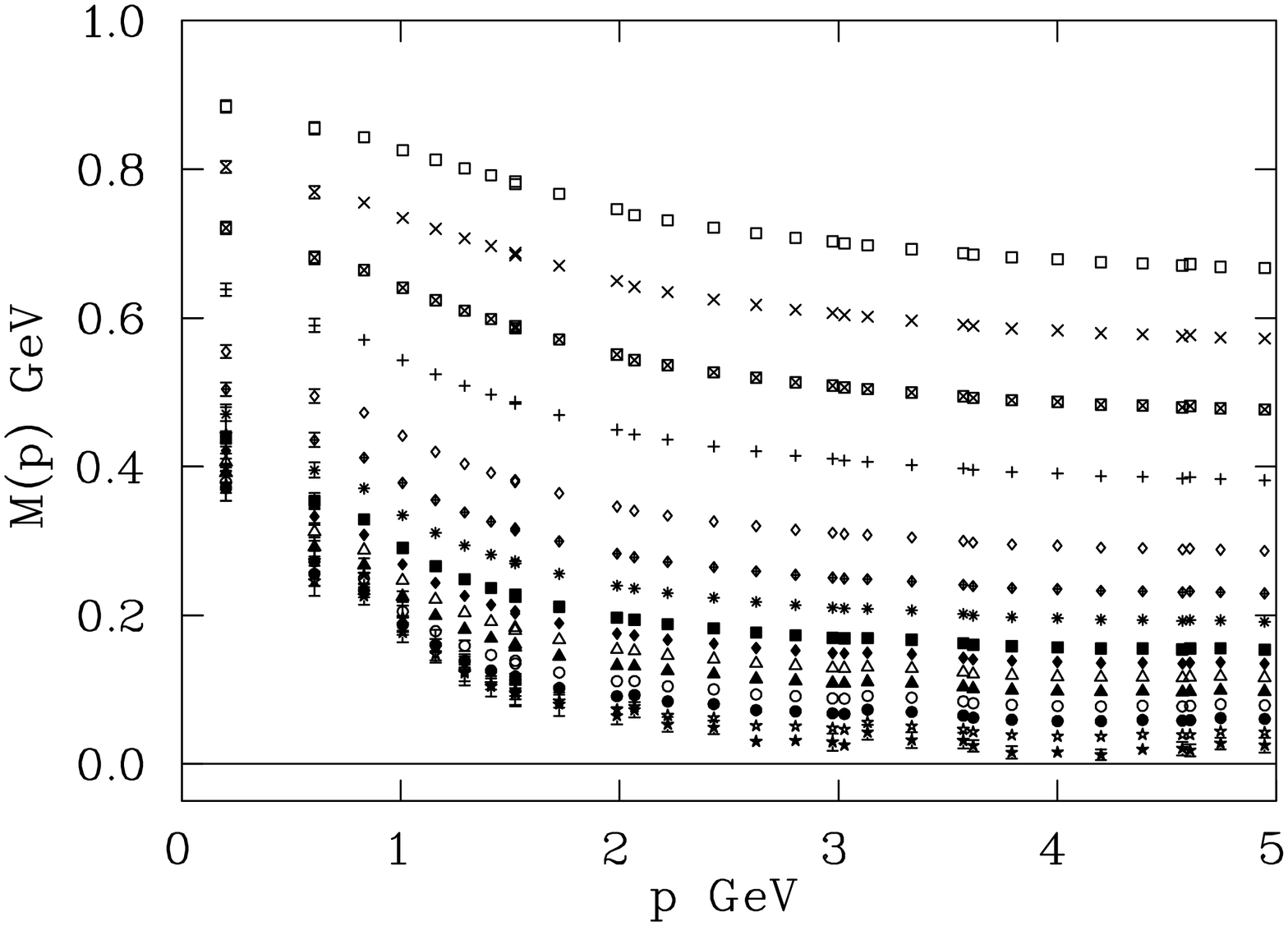}
\includegraphics[angle=90,height=0.28\textheight,width=0.45\textwidth]{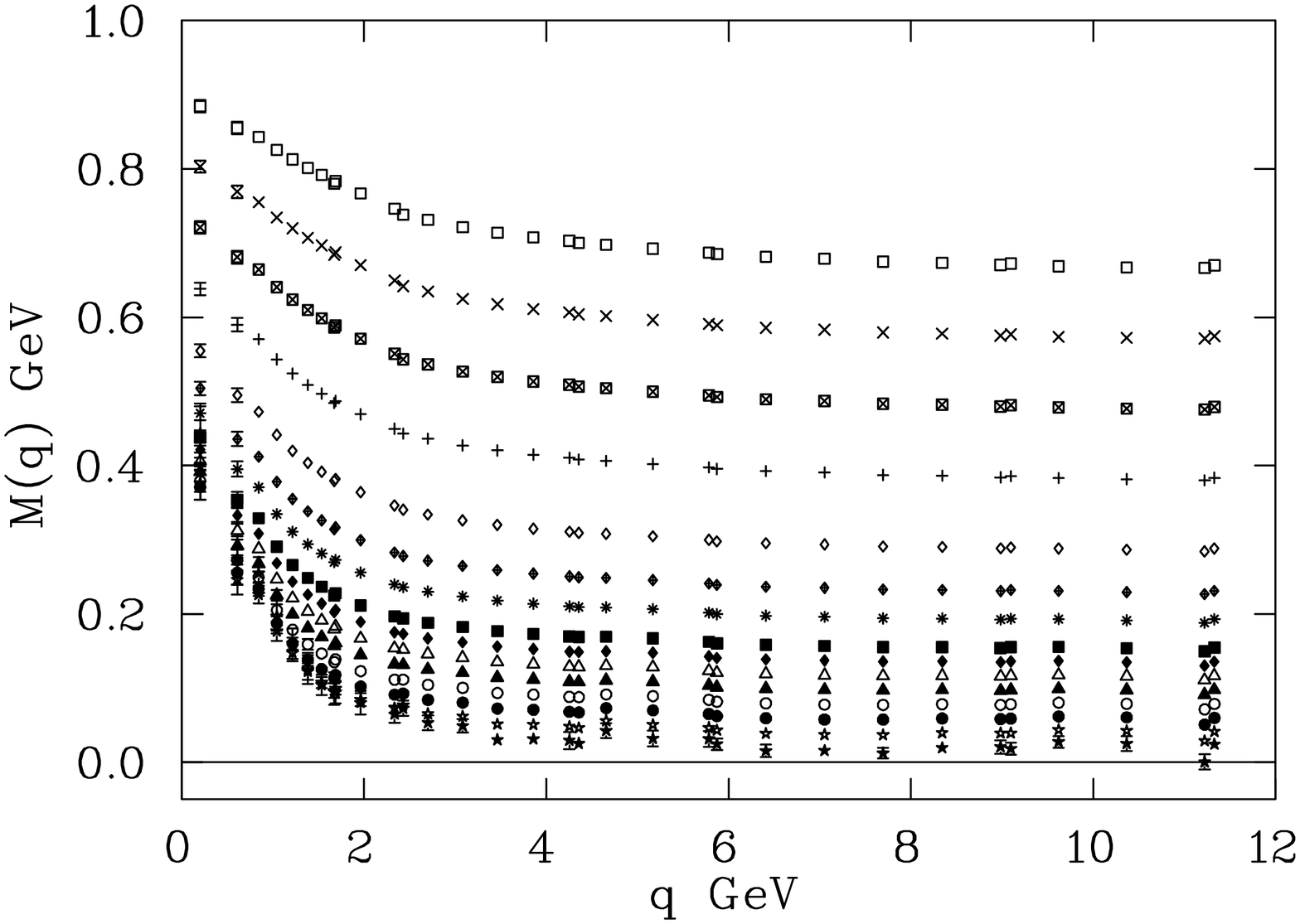}

\caption{Cylinder cut data for the FLIC Overlap mass function $M(p)$ at finite quark mass for the for the quenched lattices at $\beta=4.286$ (top), $\beta=4.6$ (middle) and $\beta=4.8$ (bottom). The lefthand plots are against the discrete lattice momentum $p$, and the righthand plots are against the kinematical lattice momentum $q.$}
\label{fig:massfunc}
\end{figure*}

\begin{figure*}[p]

\includegraphics[angle=90,height=0.28\textheight,width=0.45\textwidth]{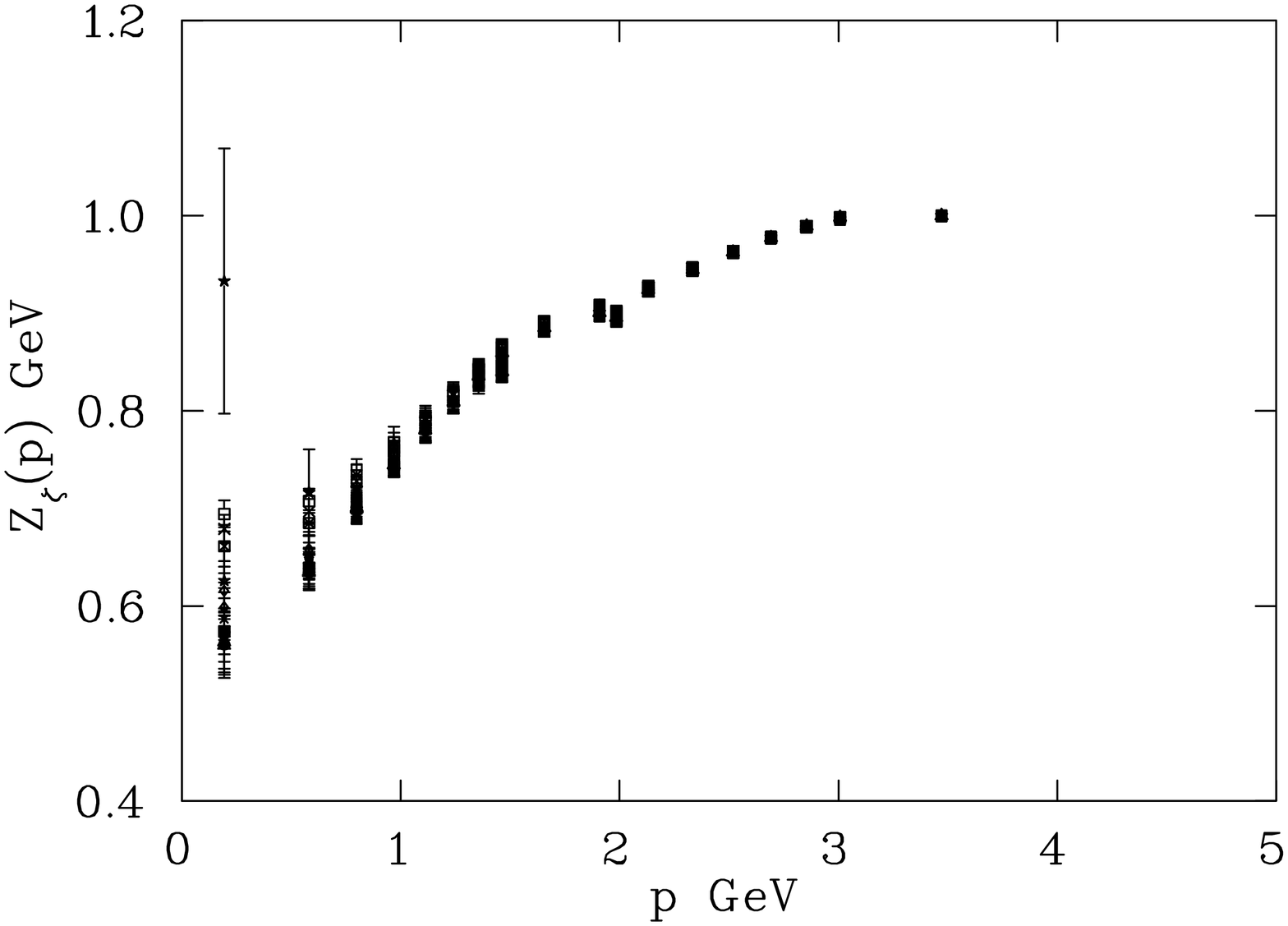}
\includegraphics[angle=90,height=0.28\textheight,width=0.45\textwidth]{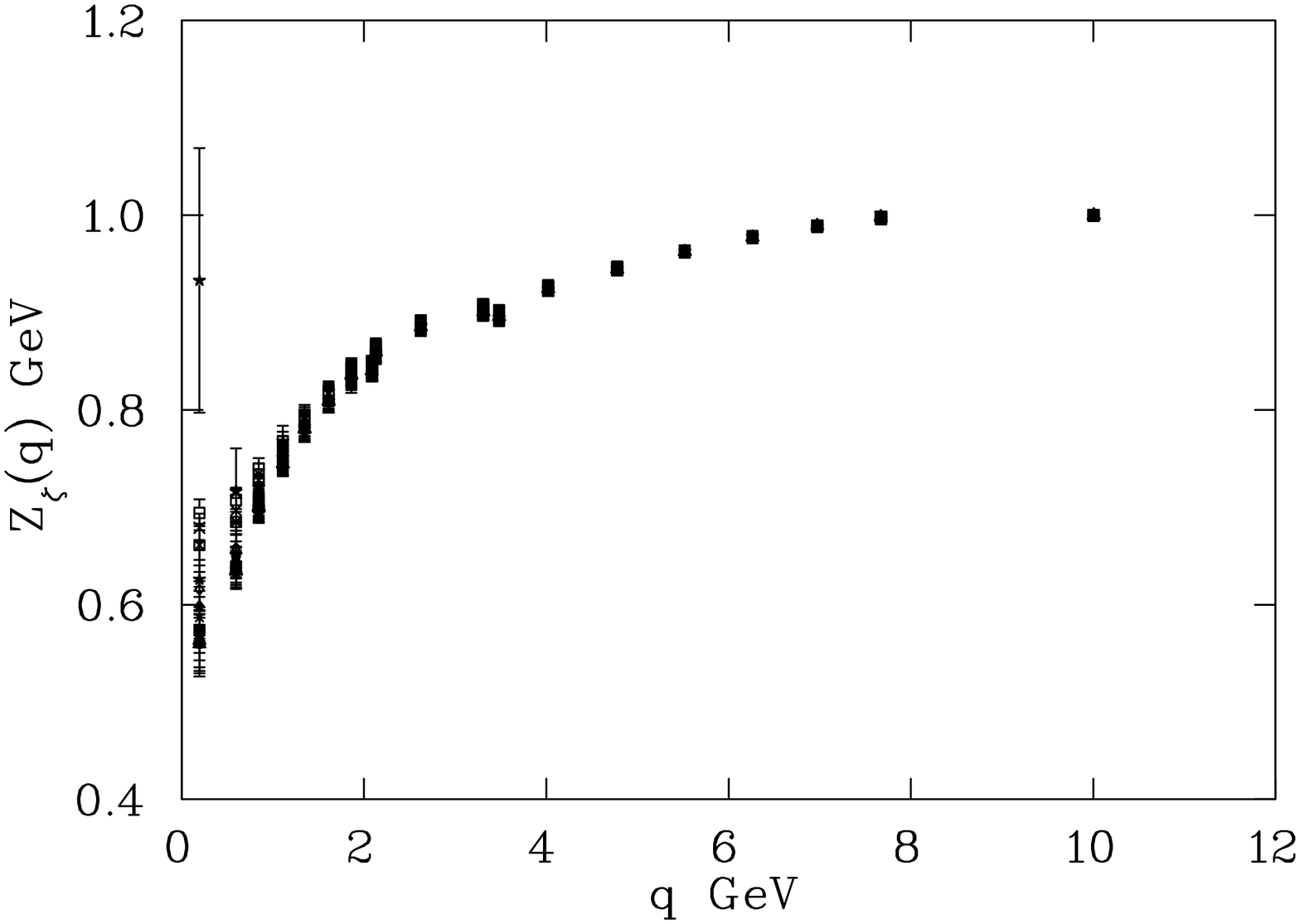}

\includegraphics[angle=90,height=0.28\textheight,width=0.45\textwidth]{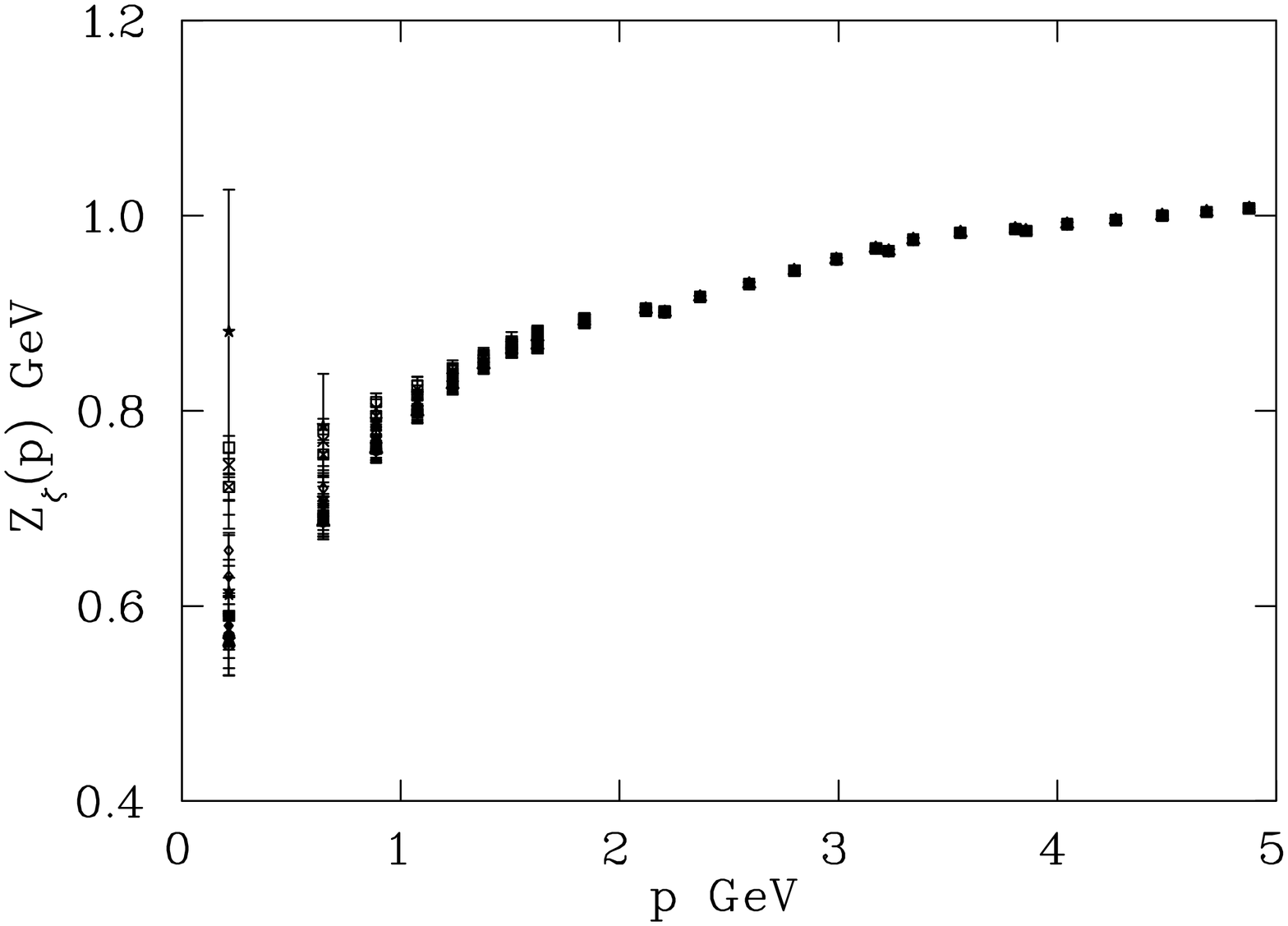}
\includegraphics[angle=90,height=0.28\textheight,width=0.45\textwidth]{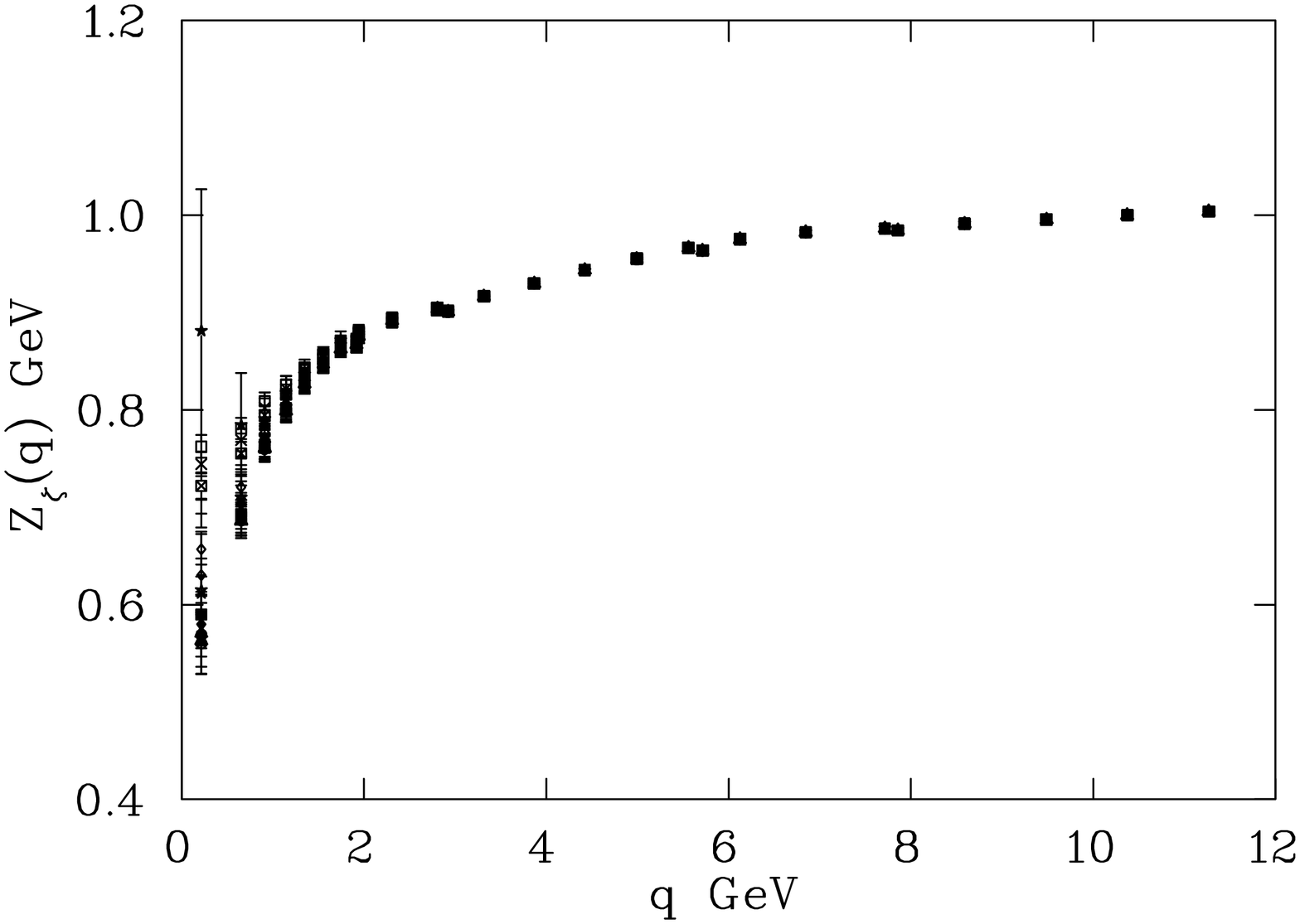}

\includegraphics[angle=90,height=0.28\textheight,width=0.45\textwidth]{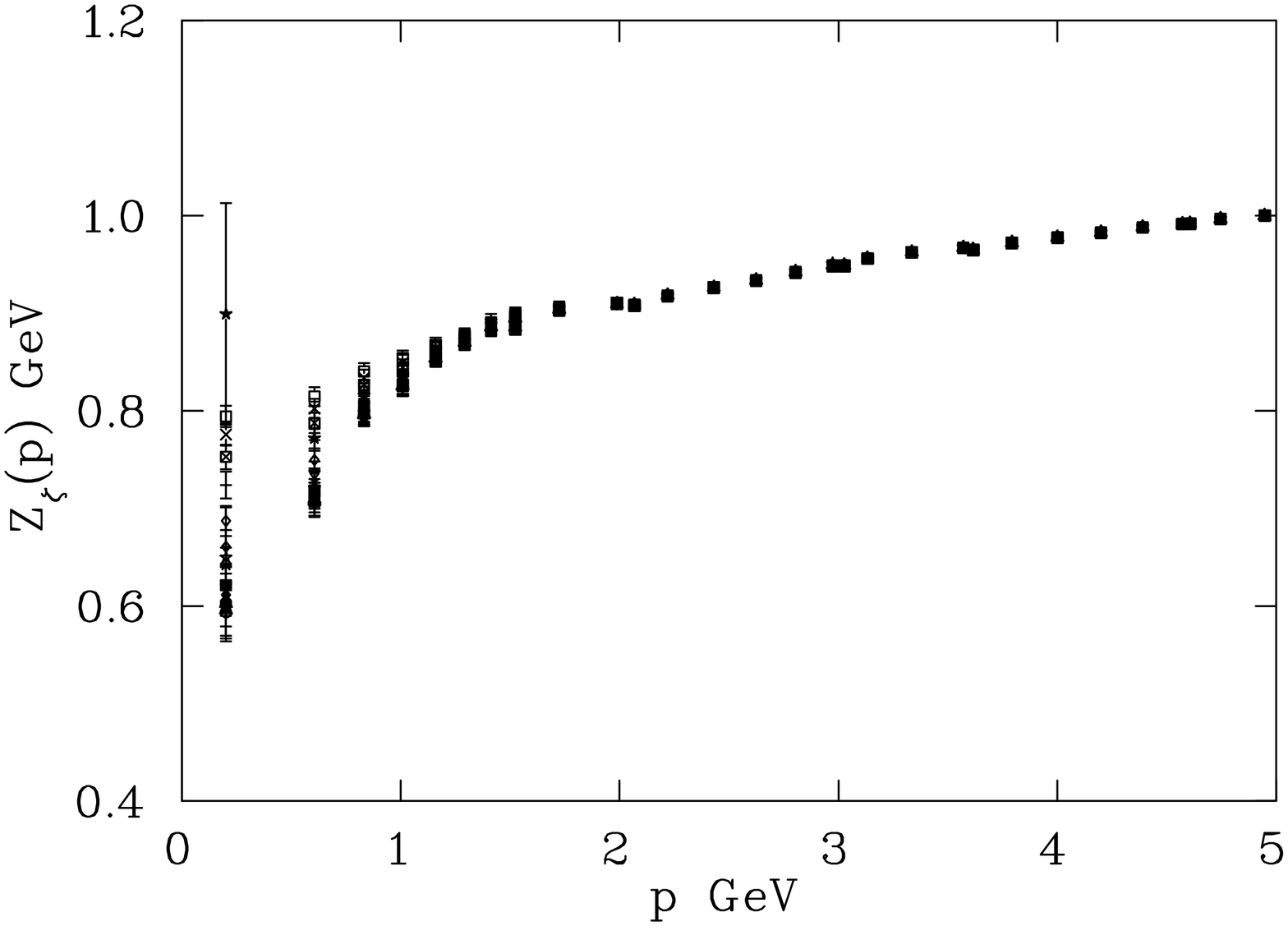}
\includegraphics[angle=90,height=0.28\textheight,width=0.45\textwidth]{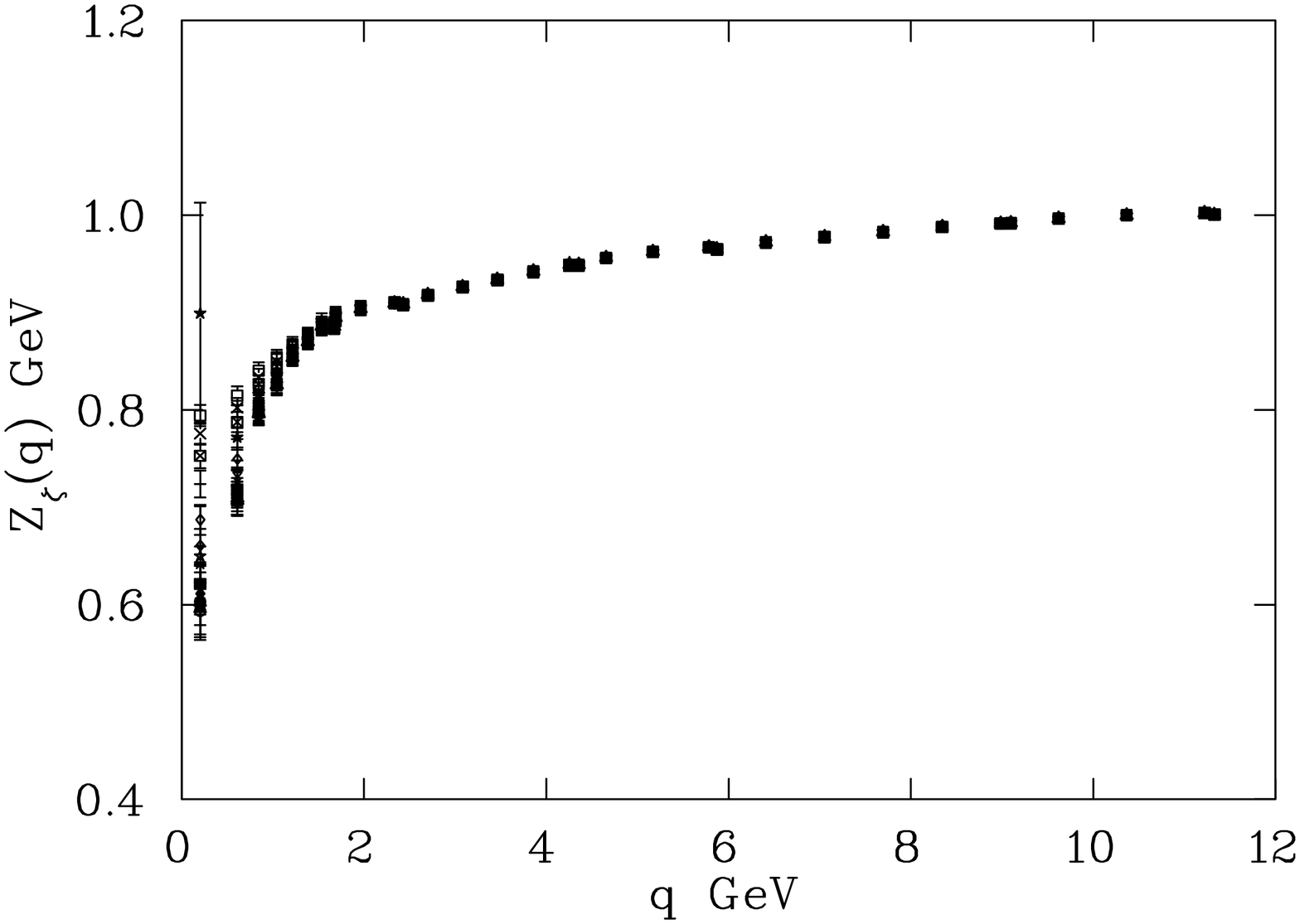}

\caption{Cylinder cut data for the FLIC Overlap renormalization function $Z_\zeta(p)$ (bottom) at finite quark mass for the quenched lattices at $\beta=4.286$ (top), $\beta=4.6$ (middle) and $\beta=4.8$ (bottom). The lefthand plots are against the discrete lattice momentum $p$, and the righthand plots are against the kinematical lattice momentum $q.$ The renormalization point for $Z_\zeta(q)$ is $\zeta=10$~GeV. The errant point in these plots is comes from the lightest quark mass.}
\label{fig:zfunc}
\end{figure*}

We now turn to the results of our investigation. The mass function $M(p)$ and the renormalisation function $Z_\zeta(p)$ are calculated on each of the lattices. A cylinder cut~\cite{overlgp} is applied to all the data, to reduce the effects of rotational symmetry violation. The cylinder cut selects momenta with $\Delta p \le \frac{2\pi}{n_i},$ where $n_i$ is the spatial extent of the lattice, and $\Delta p = |p| \sin \theta_p$ is a measure of the distance from the hypercubic diagonal $\hat{n} = \half (1,1,1,1).$ The angle $\theta_p$ is determined by $|p|\cos\theta = p\cdot\hat{n}.$ The full results for each of the lattices are displayed in Fig.~\ref{fig:massfunc} for $M(p)$ and Fig.~\ref{fig:zfunc} for $Z_\zeta(p).$  The renormalisation point $\zeta$ for $Z_\zeta(p)$ is chosen to be approximately $p=3 \text{ GeV}$ and $q=5.5\text{ GeV}$ when shown against $q.$  

\begin{figure*}[p]
\centering
\includegraphics[angle=90,height=0.28\textheight,width=0.45\textwidth]{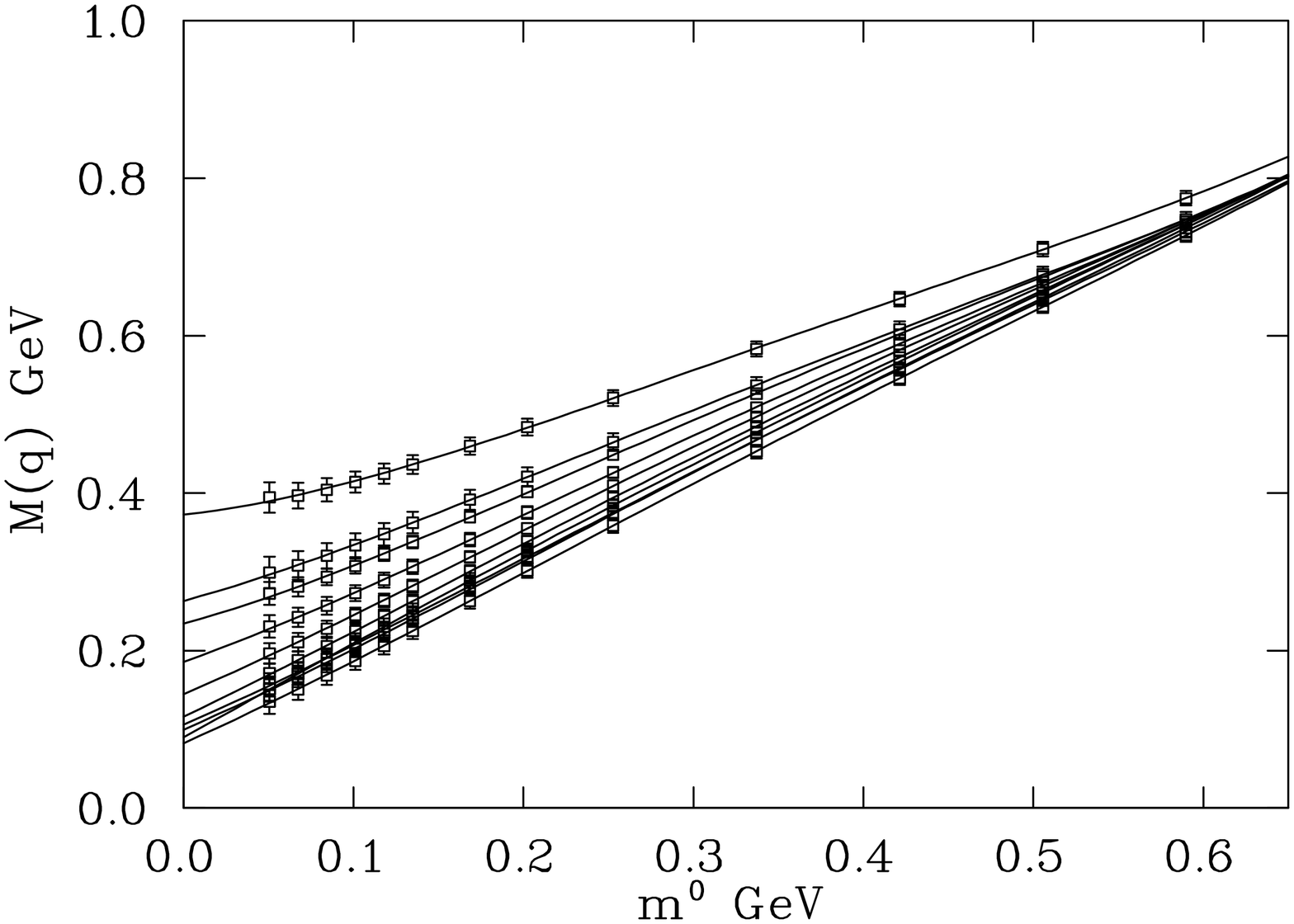}
\includegraphics[angle=90,height=0.28\textheight,width=0.45\textwidth]{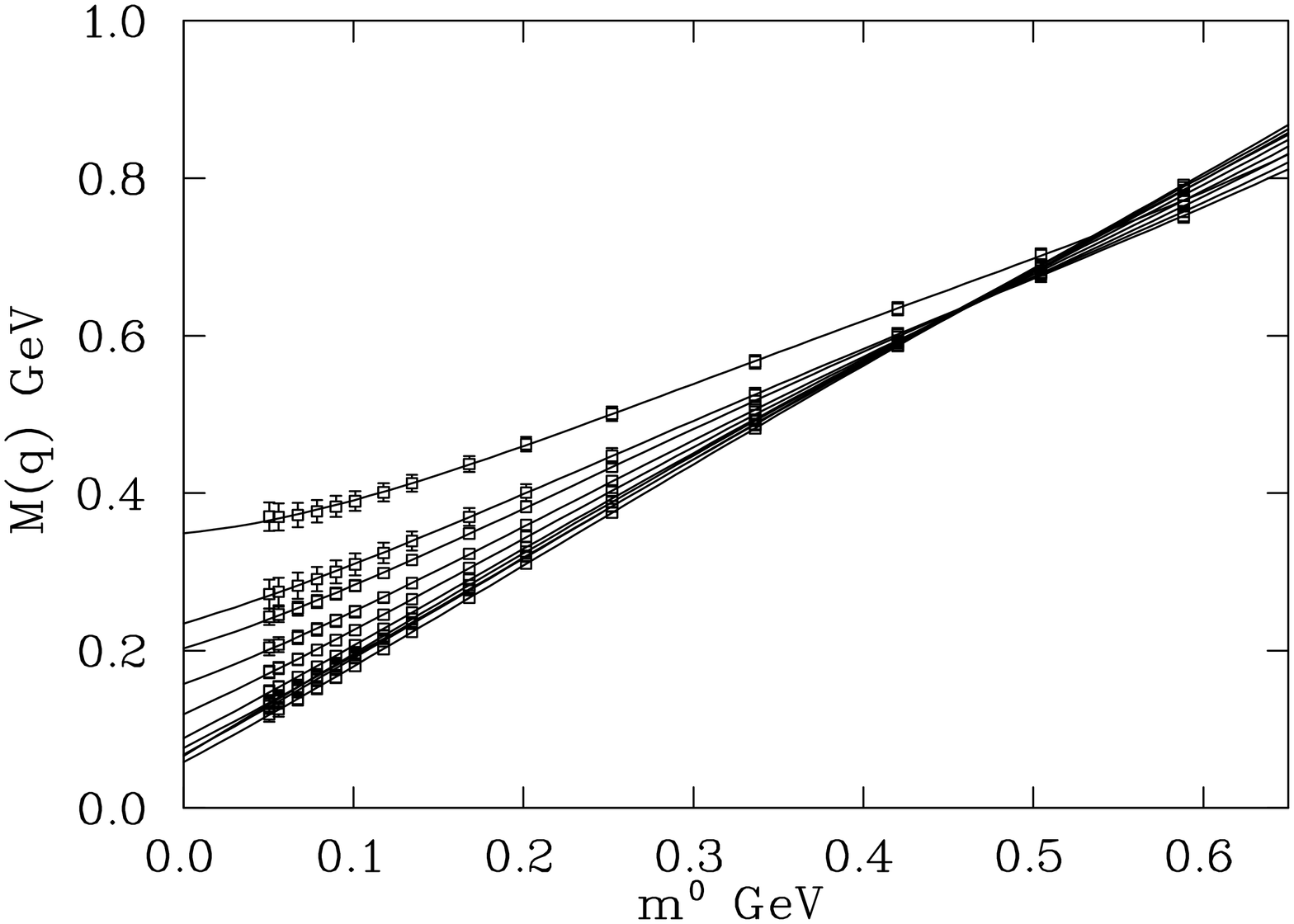}

\includegraphics[angle=90,height=0.28\textheight,width=0.45\textwidth]{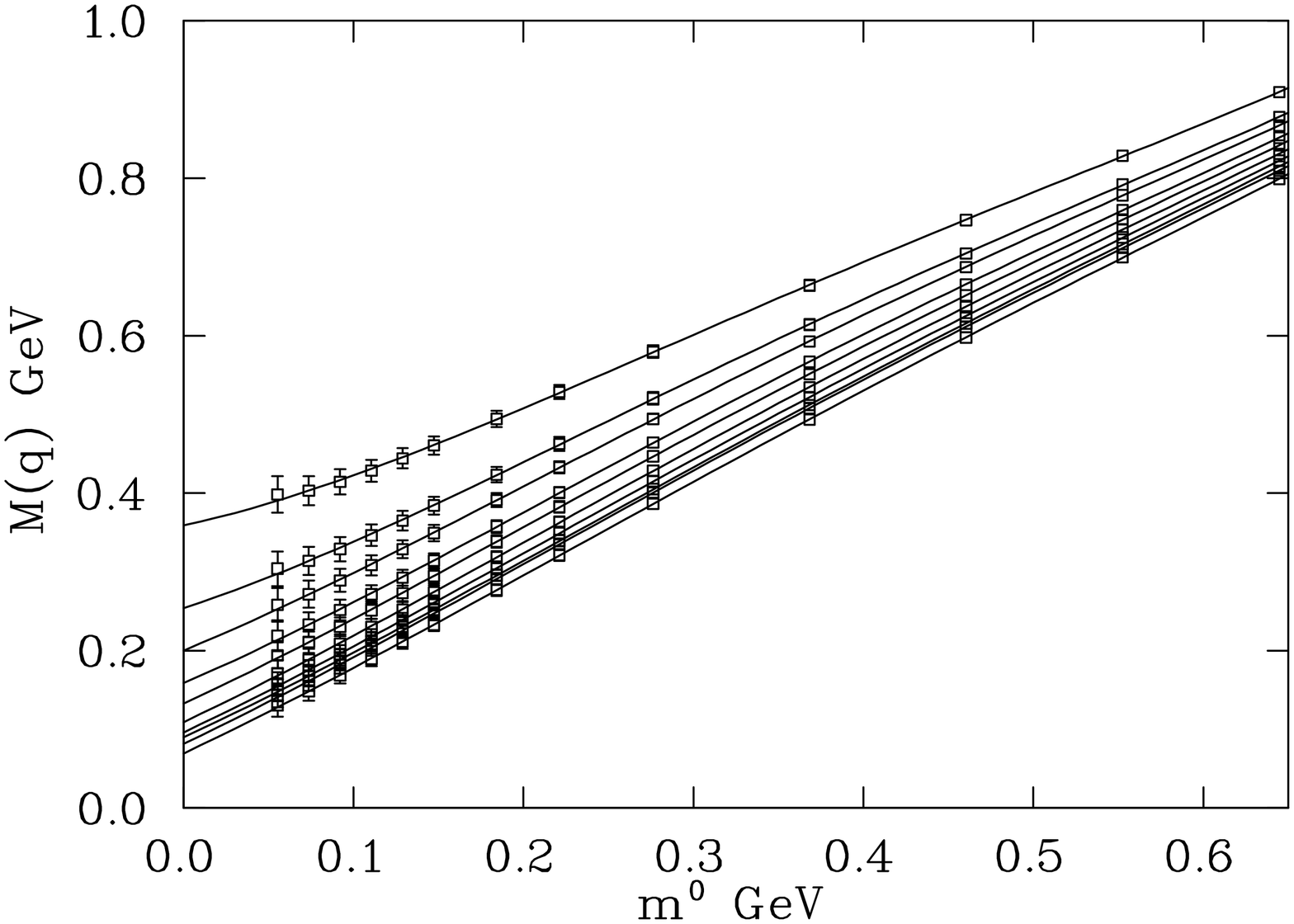}
\includegraphics[angle=90,height=0.28\textheight,width=0.45\textwidth]{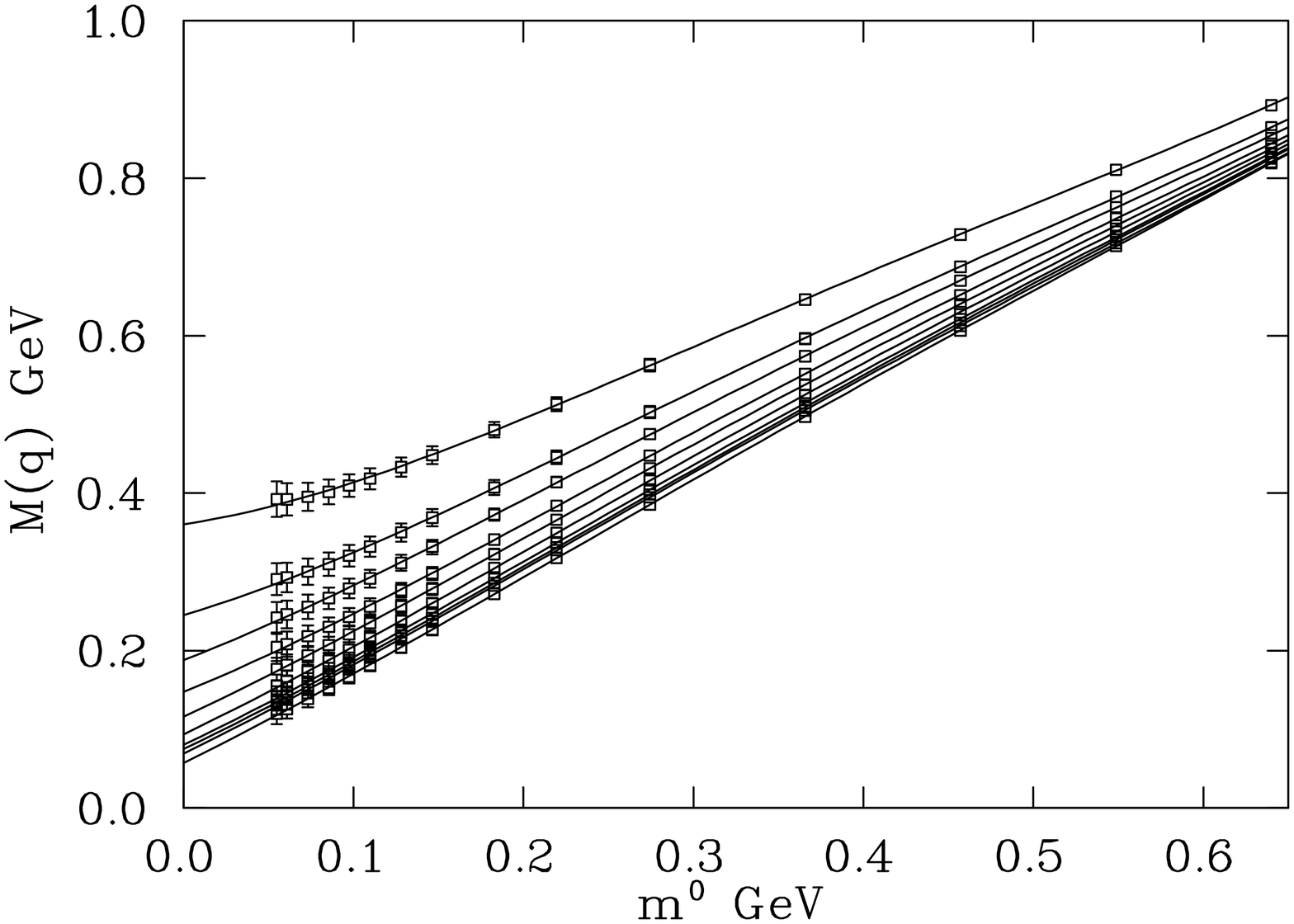}

\includegraphics[angle=90,height=0.28\textheight,width=0.45\textwidth]{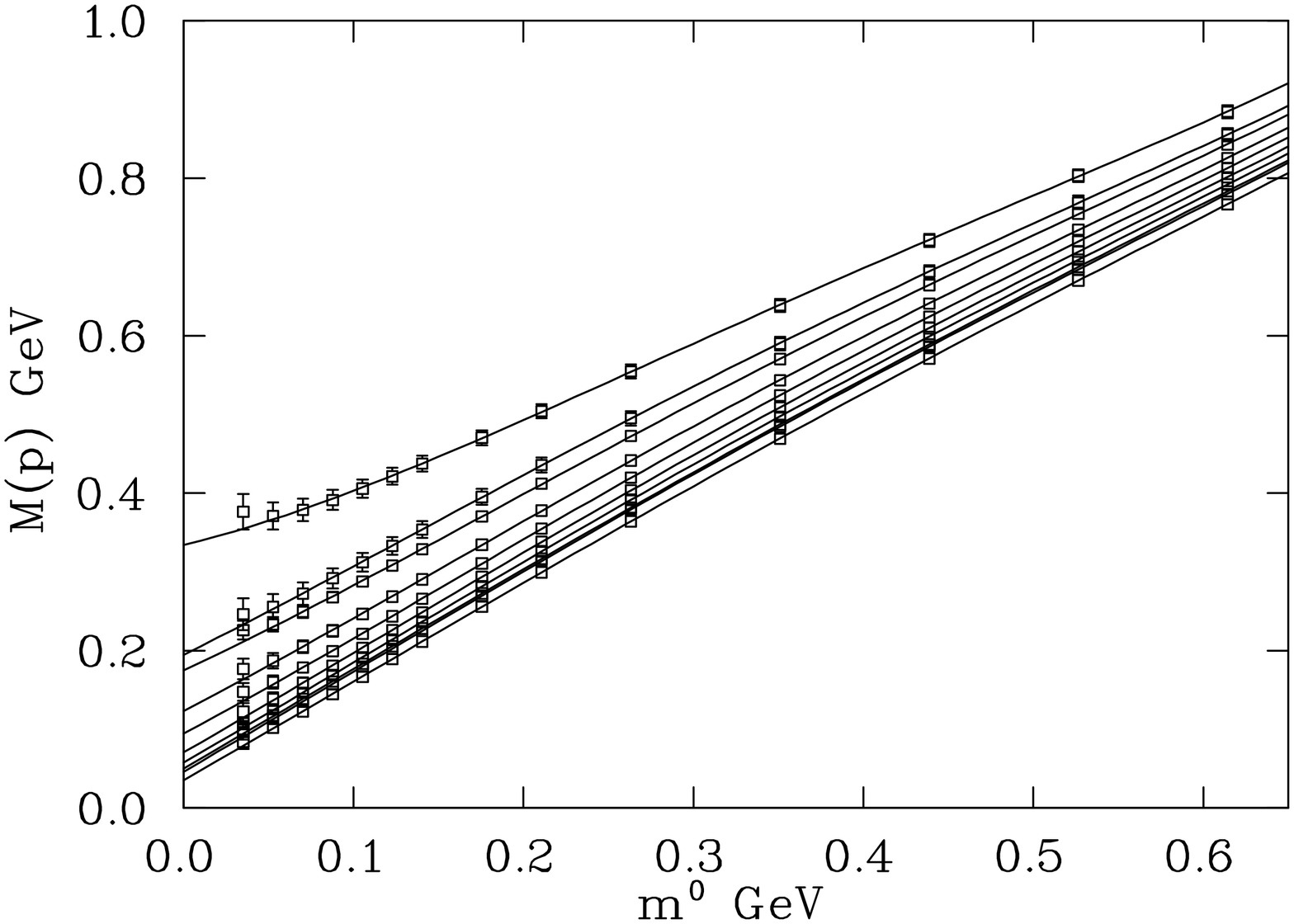}
\includegraphics[angle=90,height=0.28\textheight,width=0.45\textwidth]{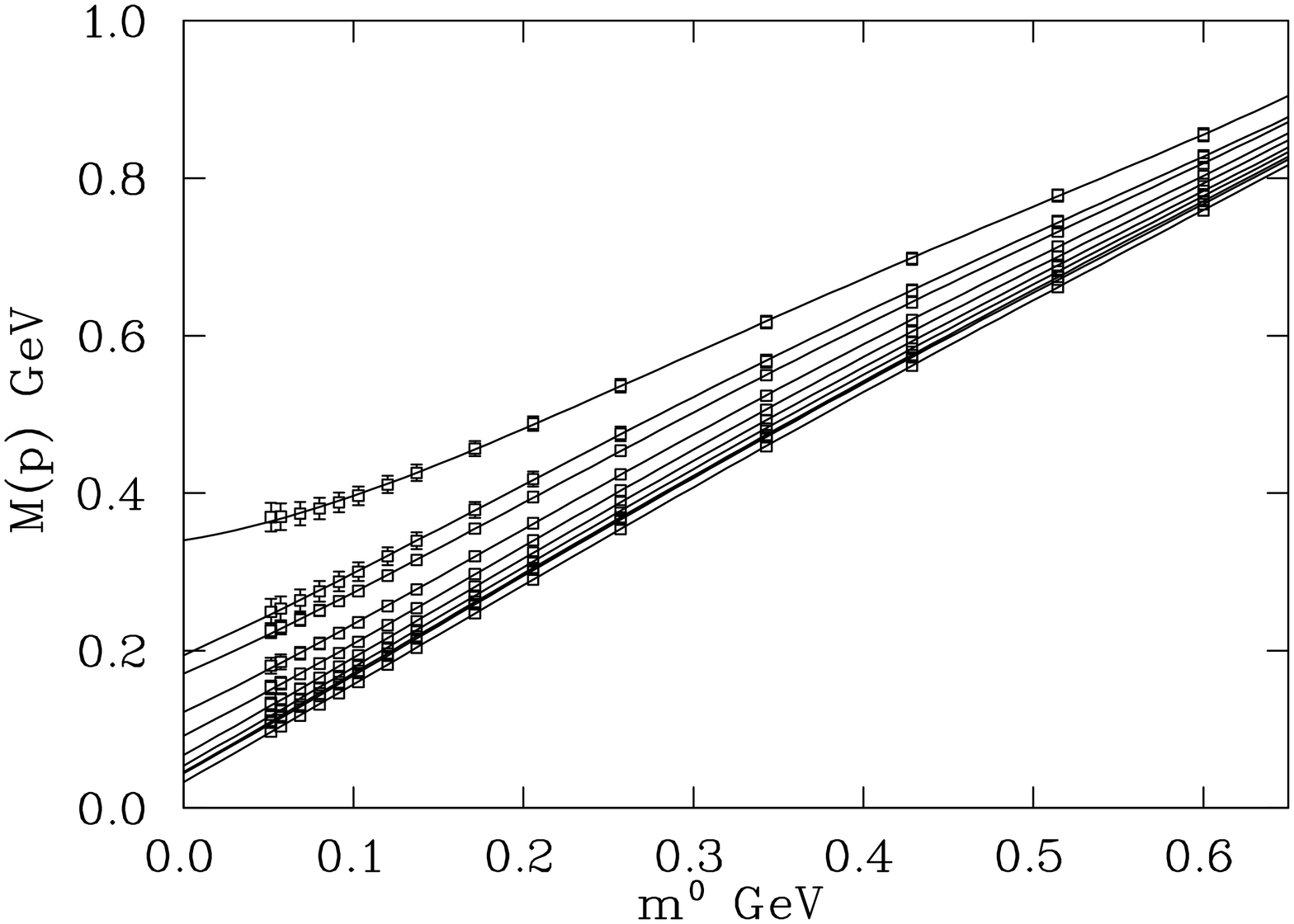}

\caption{Cylinder cut data showing the dependence of the mass function $M(p)$ for the FLIC Overlap (left) and the Wilson Overlap (right) on the bare mass $m^0$ at fixed momenta, for the lowest 10 momenta values. At small bare mass, the curves are ordered inversely to the momenta they represent, that is the smallest momenta is the topmost curve. Data is shown for for the quenched lattices at $\beta=4.286$ (top), $\beta=4.6$ (middle) and $\beta=4.8$ (top). The solid curves are the quartic fits to the data.}
\label{fig:mqchifit}
\end{figure*}

\begin{figure*}[p]
\centering
\includegraphics[angle=90,height=0.28\textheight,width=0.45\textwidth]{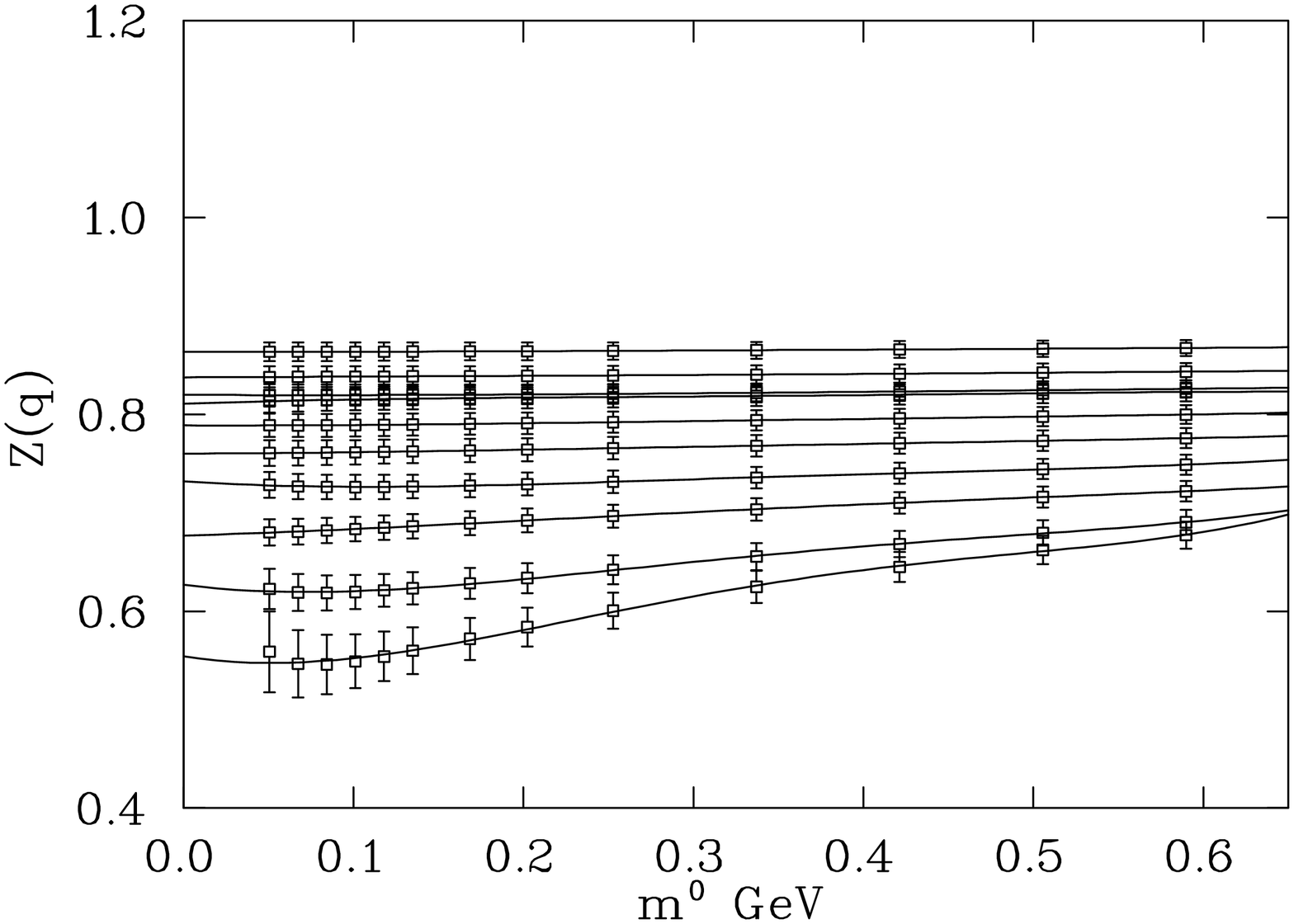}
\includegraphics[angle=90,height=0.28\textheight,width=0.45\textwidth]{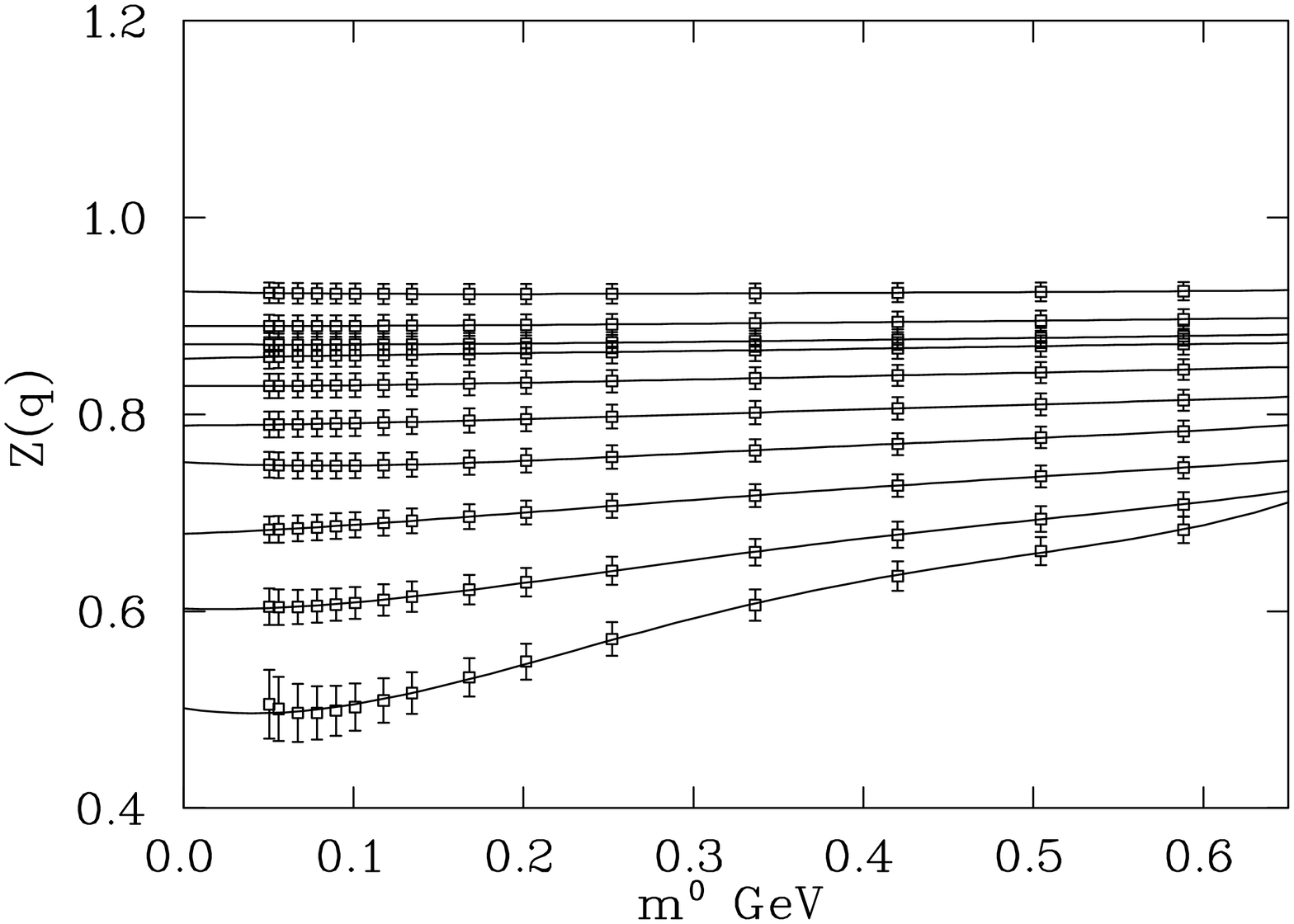}

\includegraphics[angle=90,height=0.28\textheight,width=0.45\textwidth]{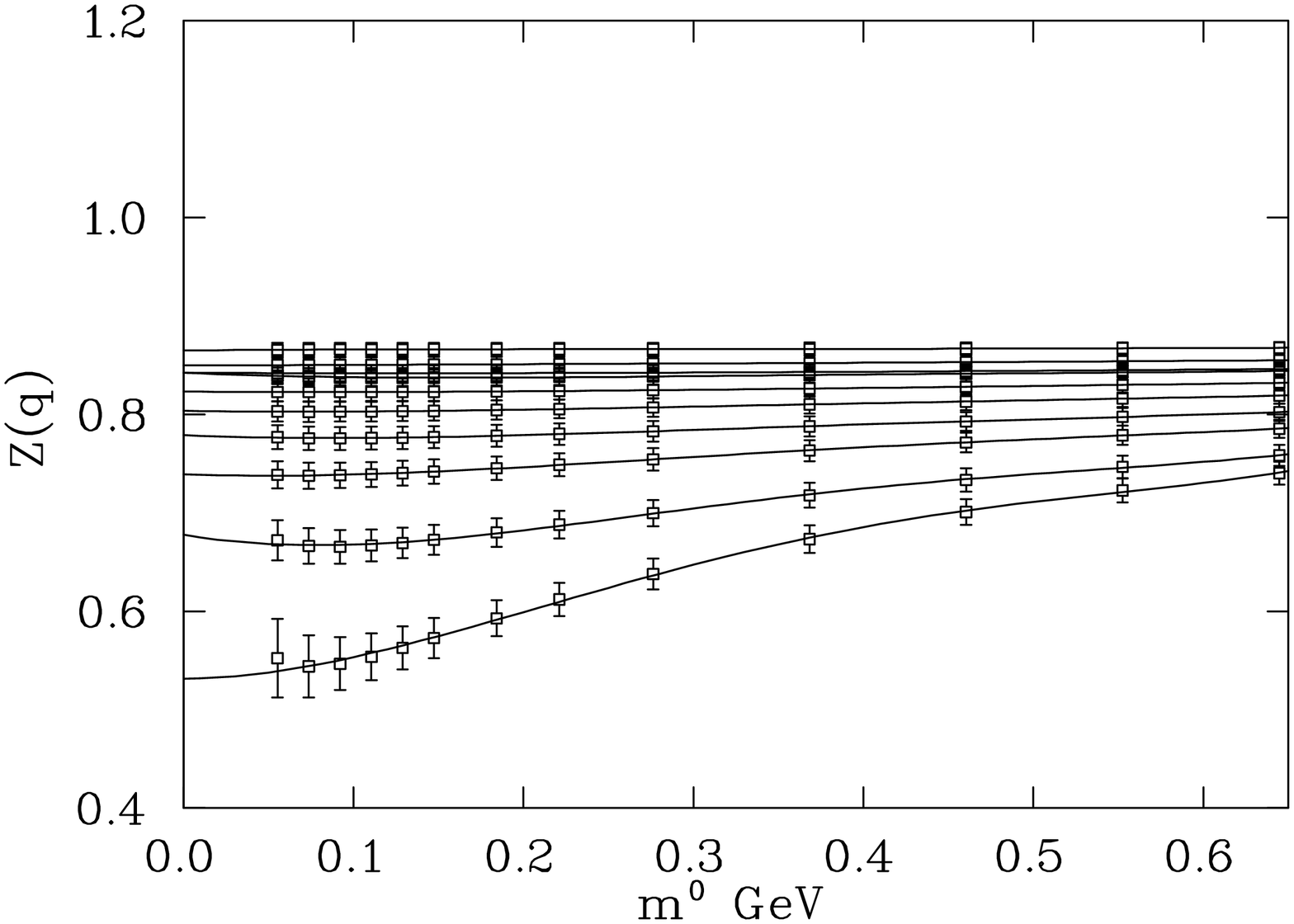}
\includegraphics[angle=90,height=0.28\textheight,width=0.45\textwidth]{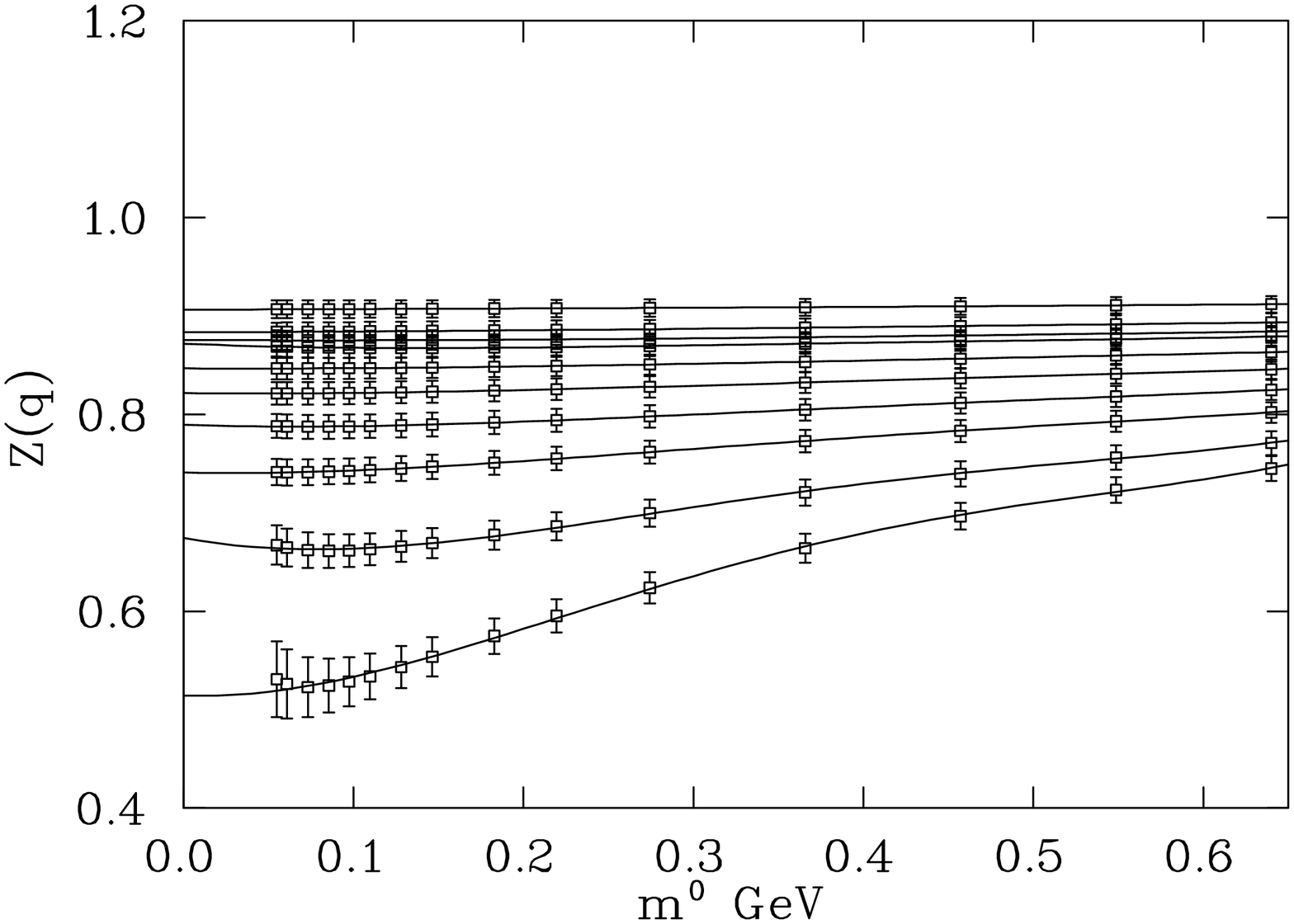}

\includegraphics[angle=90,height=0.28\textheight,width=0.45\textwidth]{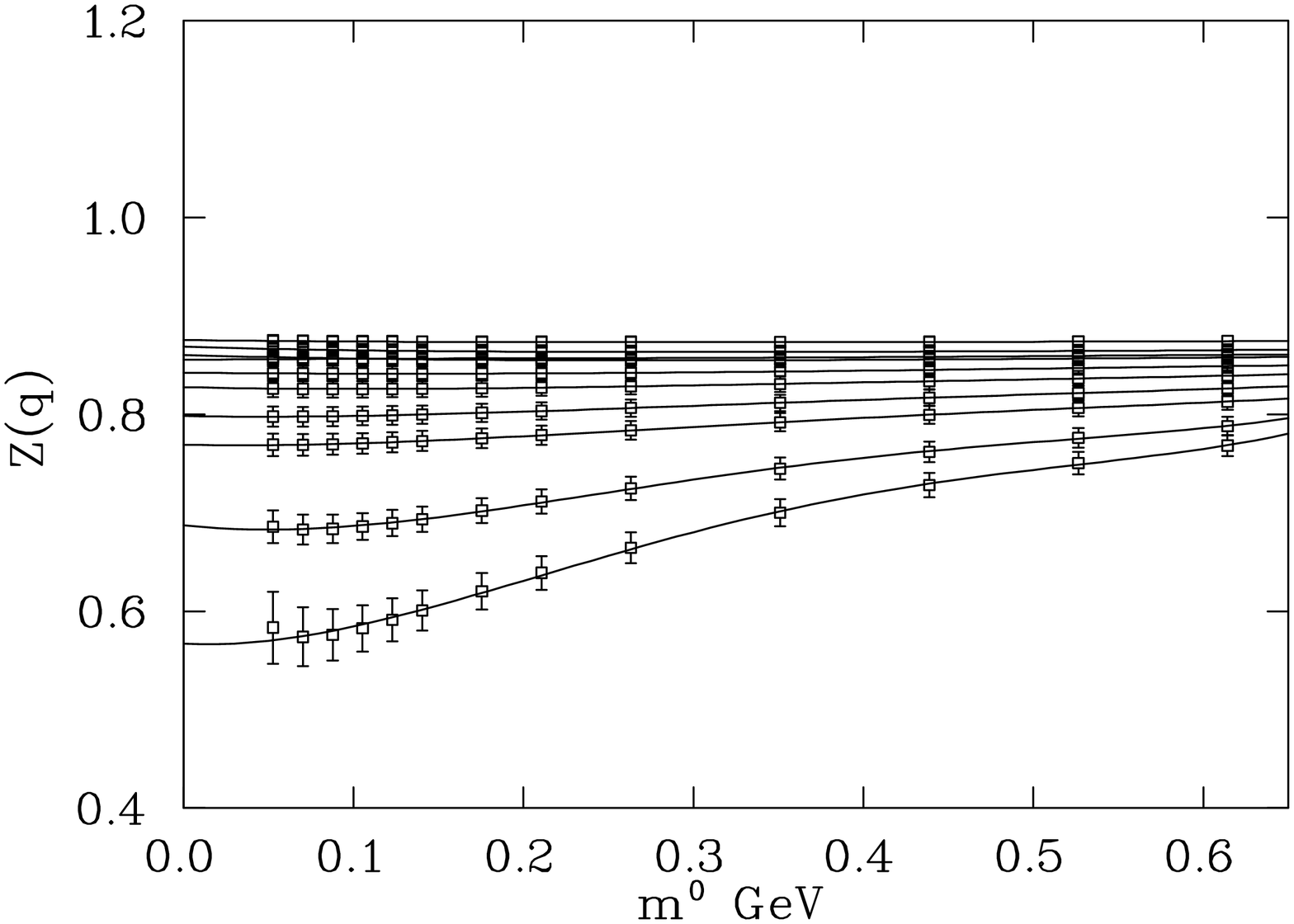}
\includegraphics[angle=90,height=0.28\textheight,width=0.45\textwidth]{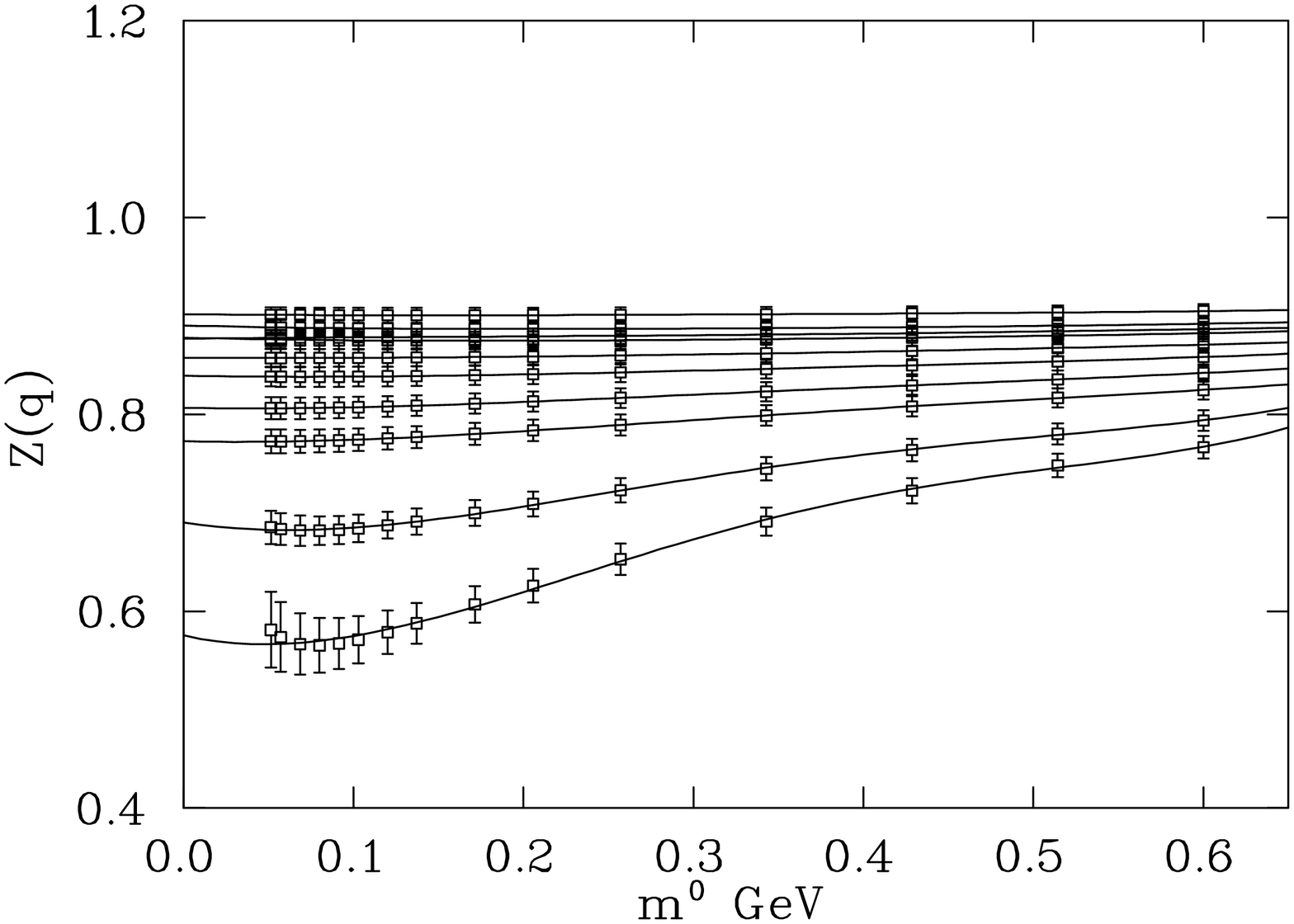}

\caption{Cylinder cut data showing the dependence of the renormalisation function $Z(p)$ for the FLIC Overlap (left) and the Wilson Overlap (right) on the bare mass $m^0$ at fixed momenta, for the lowest 10 momenta values. The curves are ordered according to the momenta they represent, that is the largest momenta is the topmost curve. Data is shown for for the quenched lattices at $\beta=4.286$ (top), $\beta=4.6$ (middle) and $\beta=4.8$ (top). The solid curves are the quartic fits to the data.}
\label{fig:zqchifit}
\end{figure*}

While the quark masses obtained from the FLIC overlap and Wilson overlap calculations are similar, they are not matched. In particular, the lightest five FLIC overlap masses are significantly lighter than the corresponding Wilson overlap masses. In order to compare the two actions, for each fixed momenta value we perform a quartic fit of both $M(p)$ and $Z(p)$ as a function of $m^0 = 2 m_{\rm w} \mu,$ (excluding the lightest two masses for the FLIC Overlap, so that the lightest bare mass included in the fit is approximately the same for both actions). The data and fits for the smallest 10 momenta on each of the lattice are shown in Fig.~\ref{fig:mqchifit} for $M(p)$ and Fig.~\ref{fig:zqchifit} for $Z(p).$  

We see that for the curves which correspond to larger $p$ the behaviour is essentially linear in the bare mass. However, for the curves which correspond to smaller $p$ we see that at heavy quark masses the behaviour is linear, while at moderate to light quark masses there is an increasing amount of curvature present. This is consistent with the behaviour of the hadron spectrum predicted by chiral perturbation theory. However, chiral perturbation theory does not give us a prediction of QCD type quantities such as $M(p).$ In this work we used a (quartic) polynomial fit function, which would not take into account any non-analytic behaviour that might appear near the chiral limit. Using a reliable formula for the extrapolation of $M(p)$ and $Z(p)$ requires further study, as we must disentangle the effects of finite volume, quenched chiral logs and genuine curvature. We note at this point that on the $\beta=4.286$ lattice the Wilson overlap masses change order at large quark mass, an indication that the FLIC overlap is superior for heavy quarks.

\begin{figure*}[p]
\centering
\includegraphics[angle=90,height=0.28\textheight,width=0.45\textwidth]{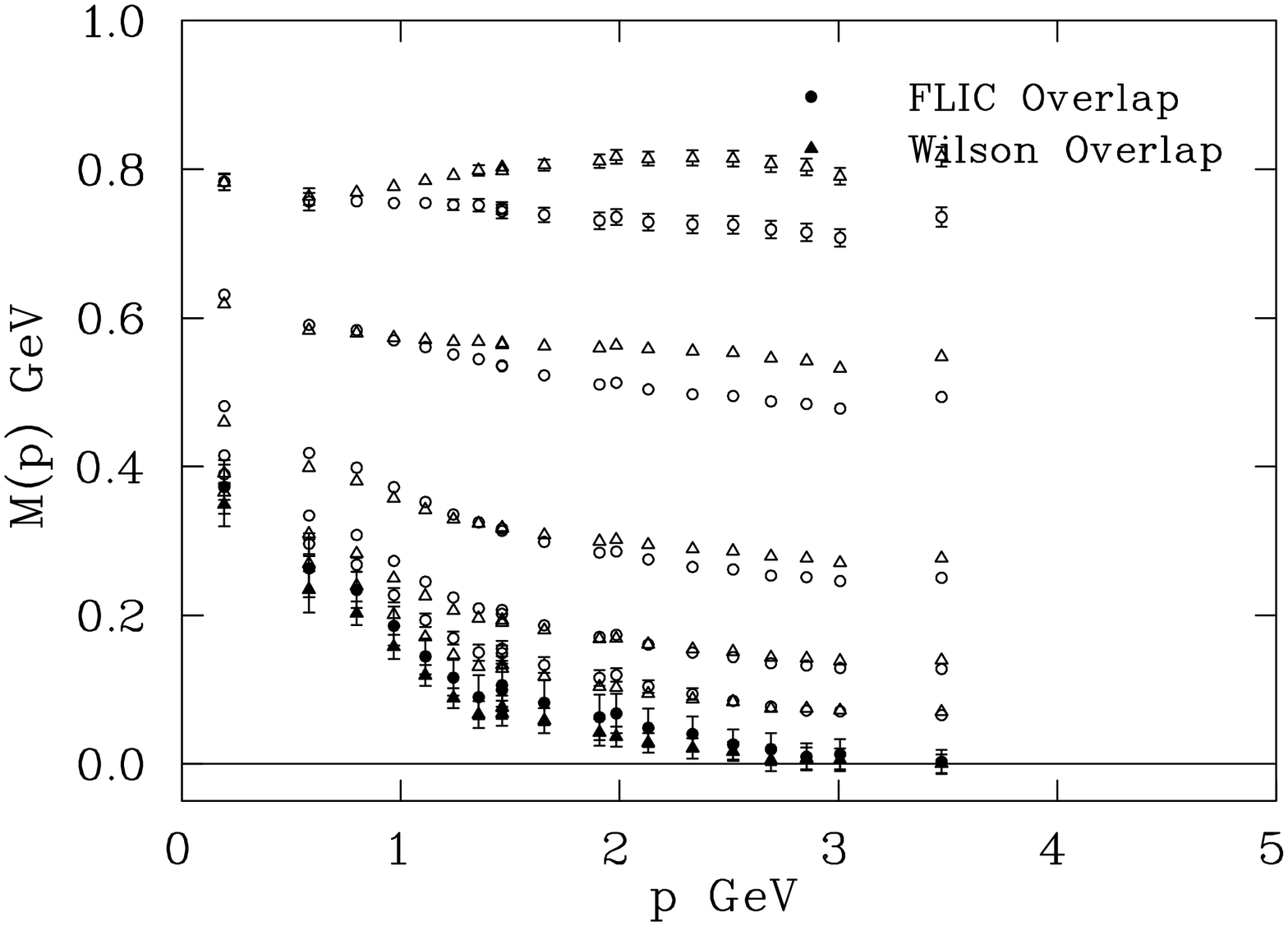}
\includegraphics[angle=90,height=0.28\textheight,width=0.45\textwidth]{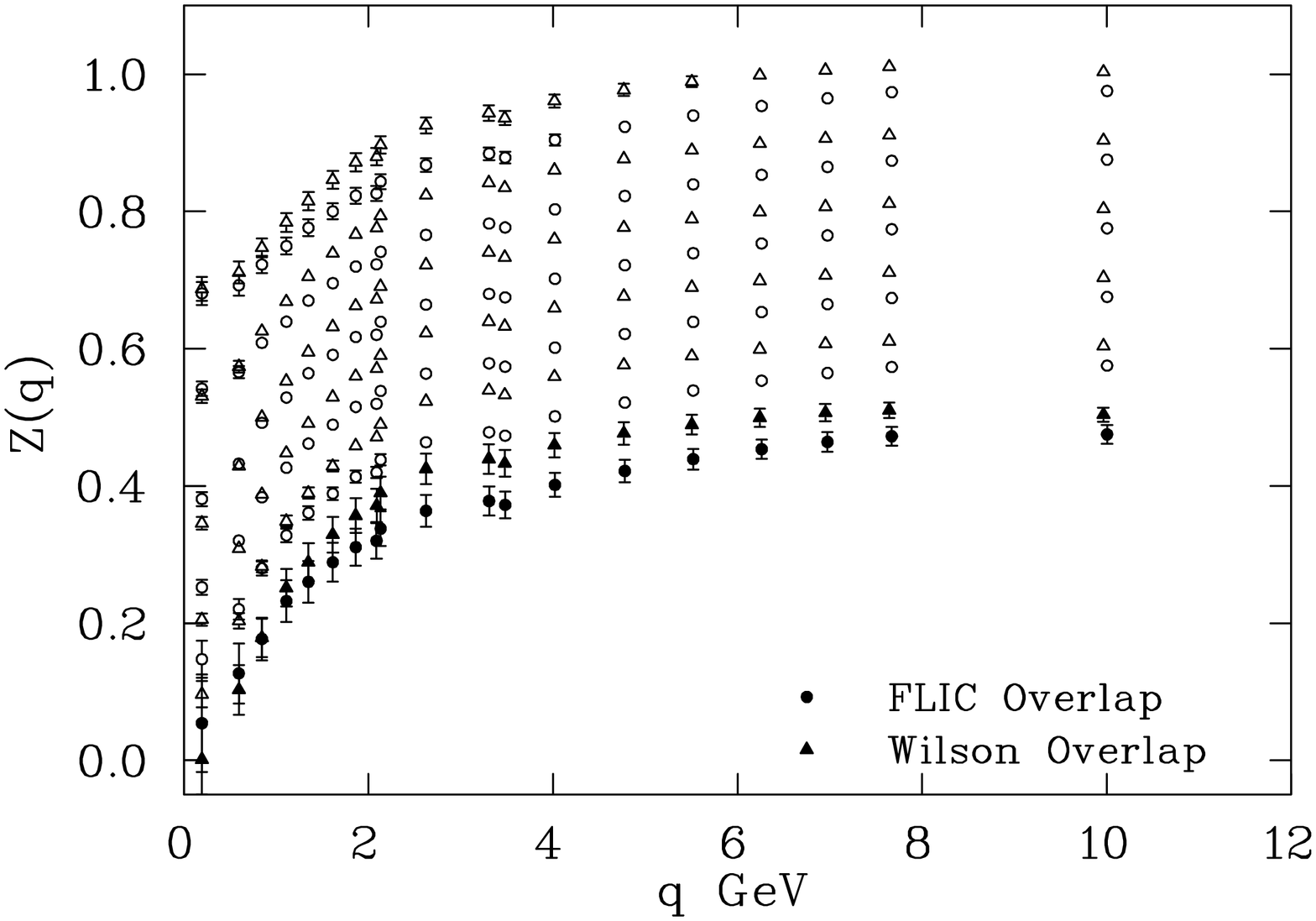}

\includegraphics[angle=90,height=0.28\textheight,width=0.45\textwidth]{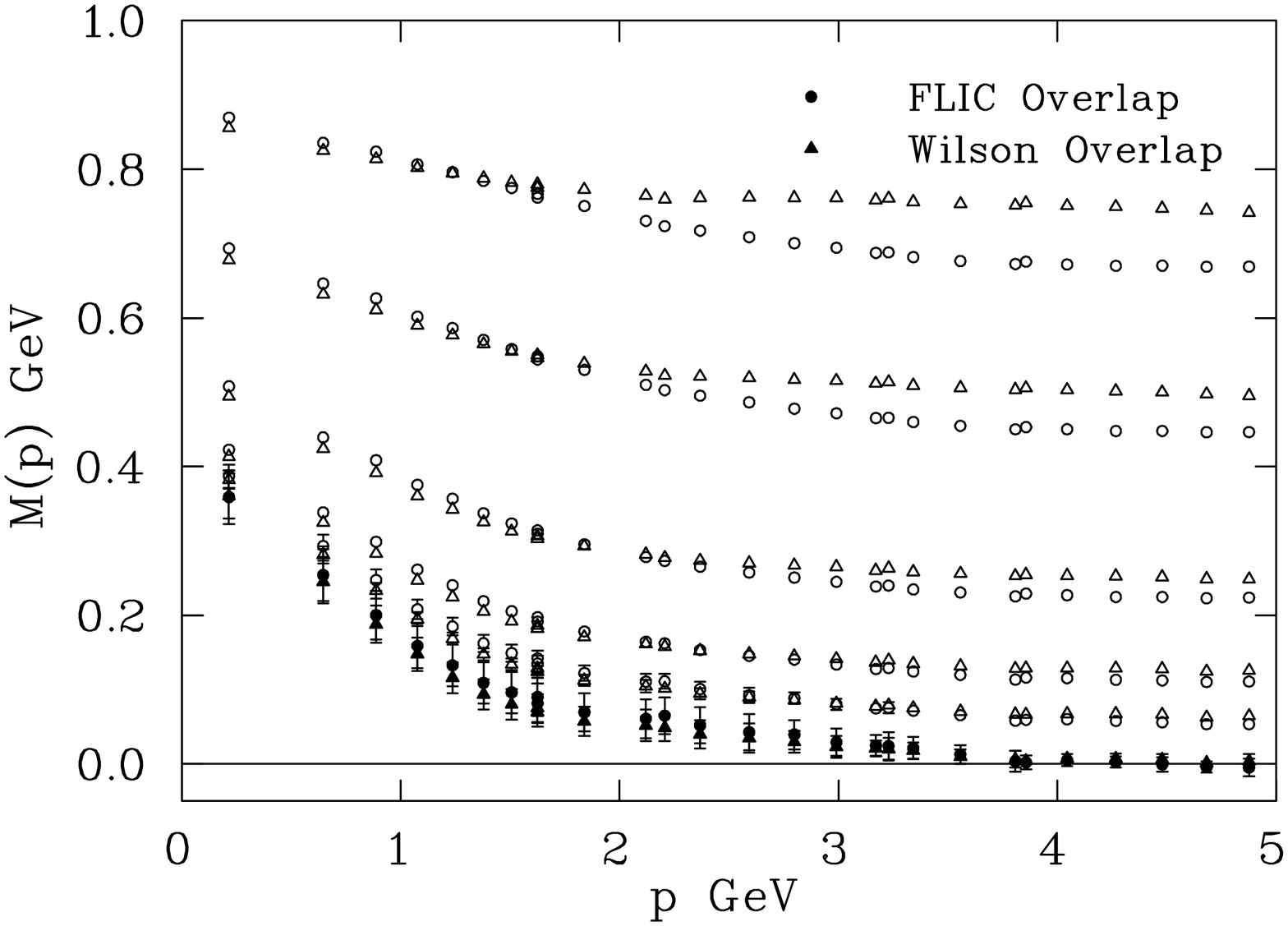}
\includegraphics[angle=90,height=0.28\textheight,width=0.45\textwidth]{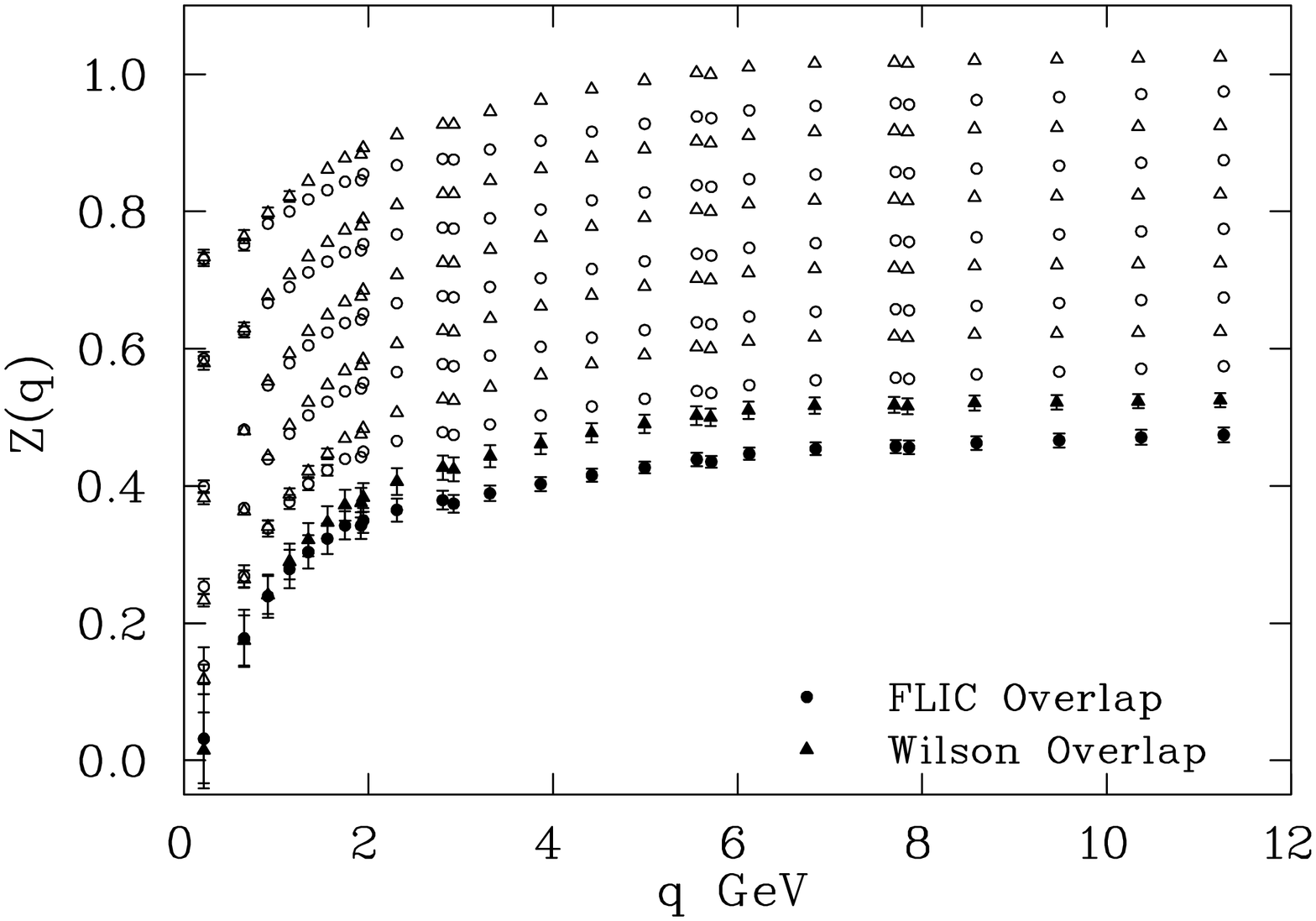}

\includegraphics[angle=90,height=0.28\textheight,width=0.45\textwidth]{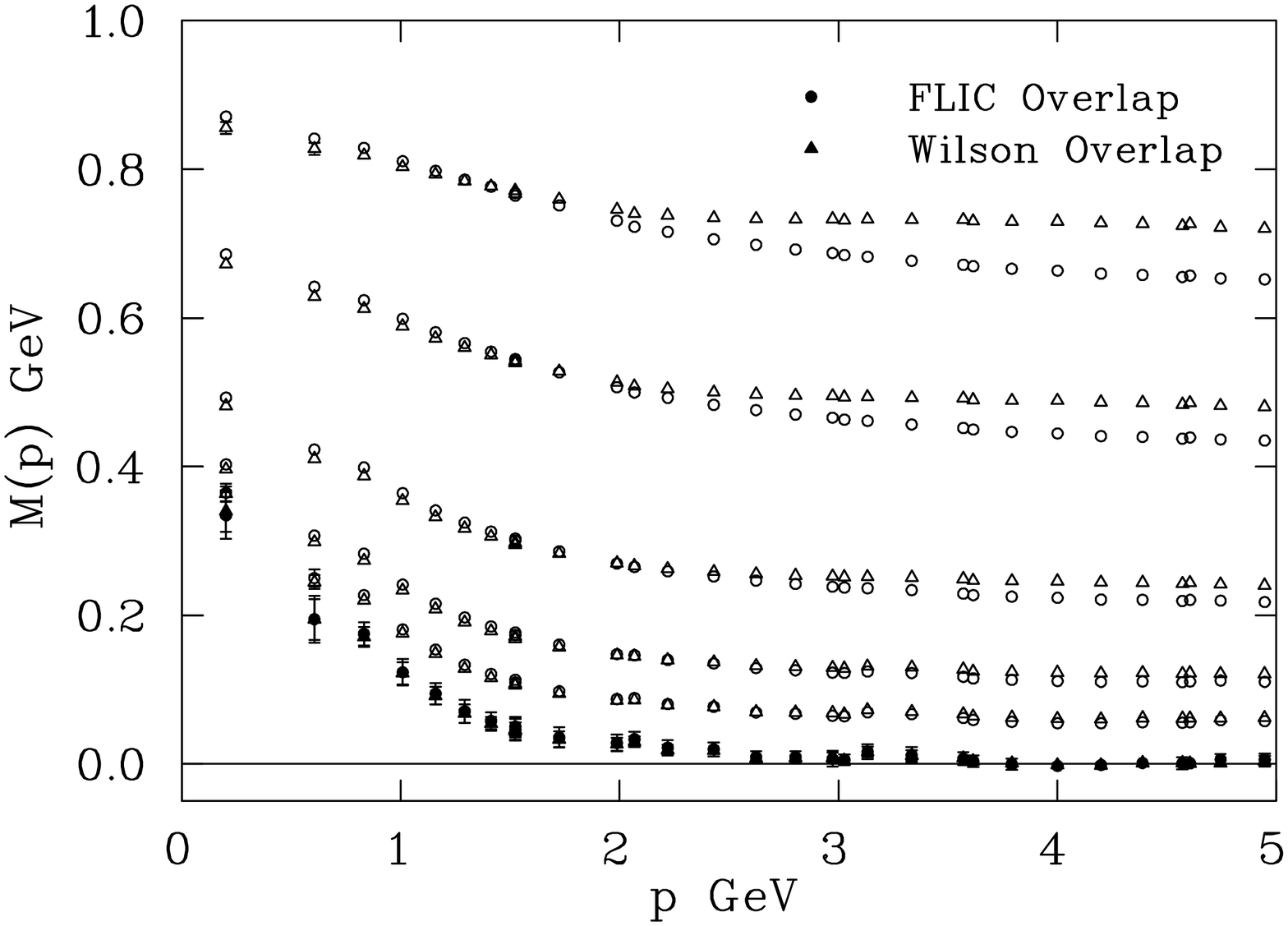}
\includegraphics[angle=90,height=0.28\textheight,width=0.45\textwidth]{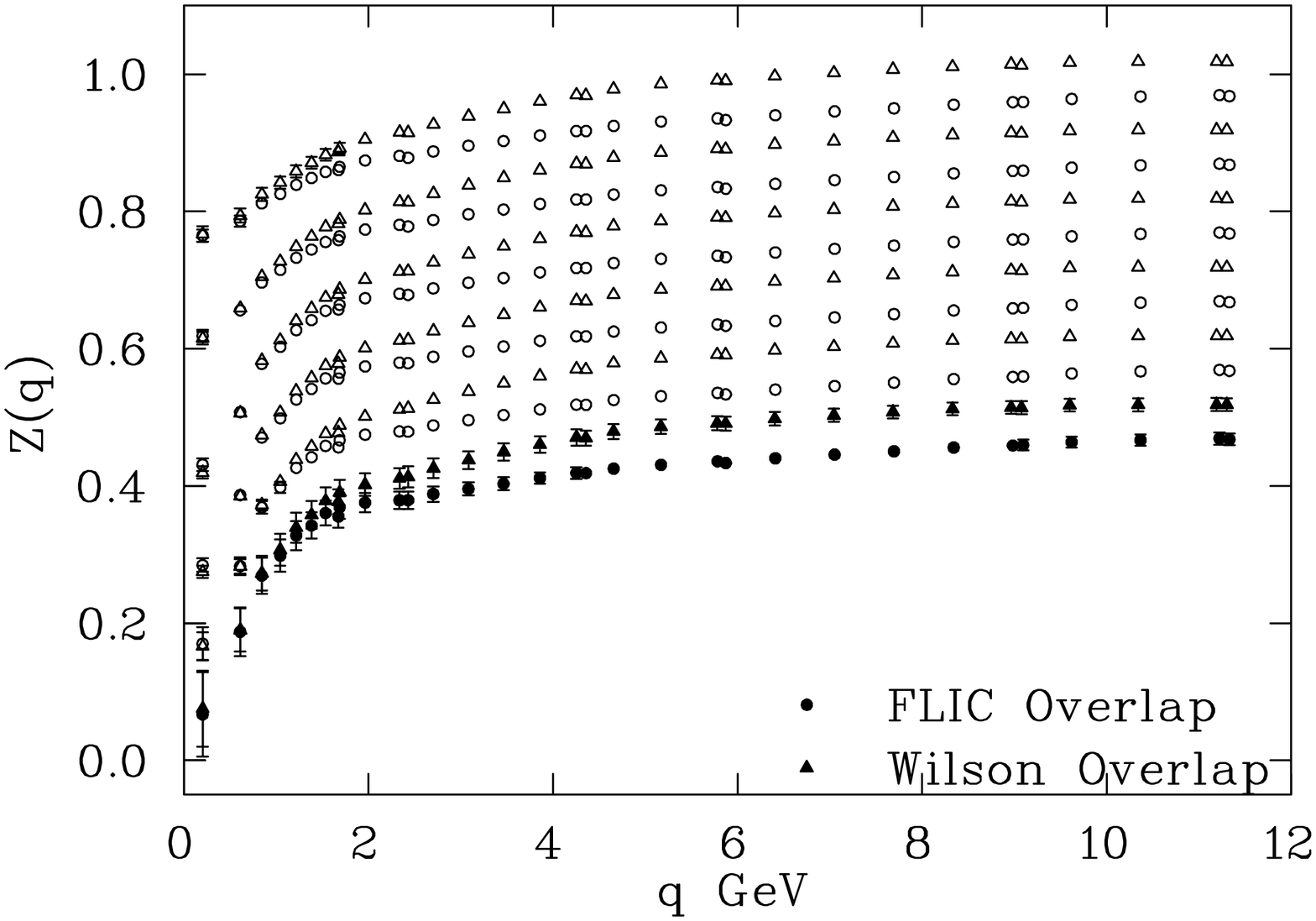}

\caption{Cylinder cut data comparing the interpolated mass function $M(p)$ (left) and renormalisation function $Z(p)$ (right), for the three quenched lattices $\beta=4.286$ (top), $\beta=4.60$ (middle), $\beta=4.80$ (bottom). The different curves from lowest to highest represent matched bare masses of $m^0 = 0.0, 0.05, 0.1, 0.2, 0.4, 0.6$ GeV. The solid points indicate the chiral limit. The different $Z(q)$ have been offset vertically for clarity.}
\label{fig:wilsonvsflic}
\end{figure*}

Having obtained the fit functions $M(p,m^0)$ and $Z(p,m^0)$ we then choose $m^0 = 0.0, 0.05, 0.1, 0.2, 0.4, 0.6$ GeV and compare the functions at these values of the bare mass. In this way we can compare the two actions with matched input masses. Results are shown in Figure \ref{fig:wilsonvsflic}, plotted against the kinematical momenta $q.$

First we examine the mass function, plotted vs $p$ (as in~\cite{overlgp3}). We see that the actions agree well in the infrared, and are both flat in the ultraviolet. At the lighter quark masses the two actions agree across the whole momentum spectrum, indicating the renormalisation of the bare mass is similar for the two actions. This is further supported by comparing the infrared behavior of the mass functions for the two actions at all quark masses. 

For the heavier quark masses different asymptotic values of the quark mass are observed. The difference becomes more pronounced as the quark mass increases. What this shows is that with a matched input bare mass, the low energy properties of the two lattice actions are the same. That is, where the lattice approximation is good, the actions agree. There are well known difficulties associated with putting heavy quarks on the lattice, as the Compton wavelength of the quarks becomes small. It is in this area that we are likely to see sensitivity to the discretisation errors, and this is shown by the data, particularly in the shape of the momentum dependence. In the very infrared the two actions agree for all masses, but at heavy quark mass as the momenta increases they begin to differ, with the coarser lattice showing a more rapid divergence (as expected). Given the relatively strange behaviour of the Wilson overlap mass function on the coarsest lattice at the heaviest masses, we can conclude that the FLIC overlap inherits the benefits of the improved ultraviolet behaviour of the FLIC kernel, due to the APE smearing of the irrelevant operators.

The comparison of the renormalisation function (which we plot vs $q$) between the two actions leads to similar conclusions as illustrated
 in Fig.~\ref{fig:wilsonvsflic}. Two things should be noted however. The first is that the two actions disagree in the ultraviolet regardless of the quark mass. This suggests that $Z(p)$ is more sensitive to the cutoff than $M(p).$ Secondly, we note that we are comparing $Z(p)$ and not $Z_\zeta(p).$ There is a good reason for this, as by fixing the behaviour of $Z_\zeta(p)$ at high momenta and then comparing the two actions, we would have been led to believe (spuriously) that the actions agreed in the ultraviolet but not in the infrared. This also tells us that if we choose $\zeta$ in the ultraviolet, then $Z_2(\zeta,a)$ is different for the two actions.

\begin{figure*}[t]
\centering
\includegraphics[angle=90,height=0.28\textheight,width=0.45\textwidth]{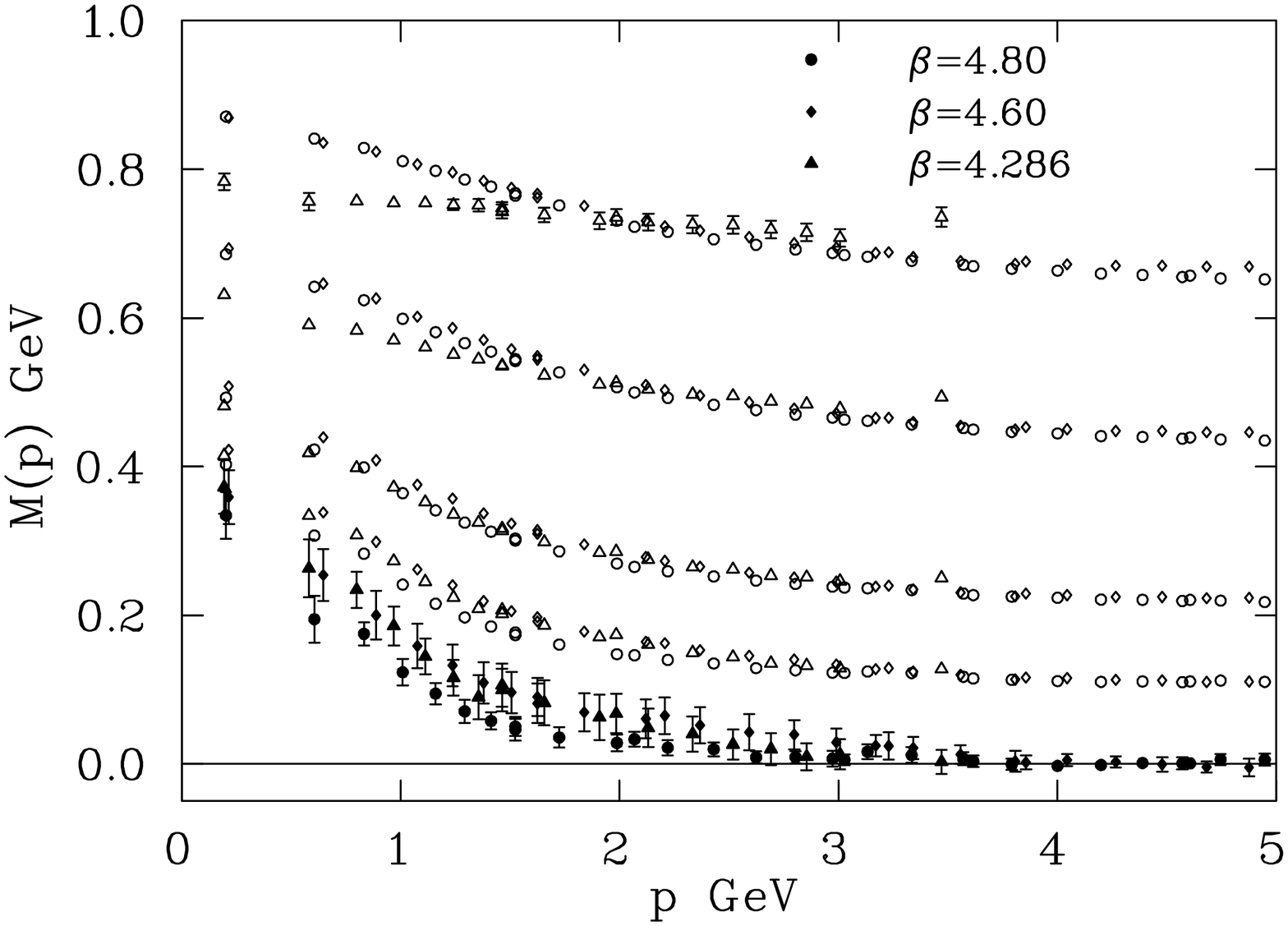}
\includegraphics[angle=90,height=0.28\textheight,width=0.45\textwidth]{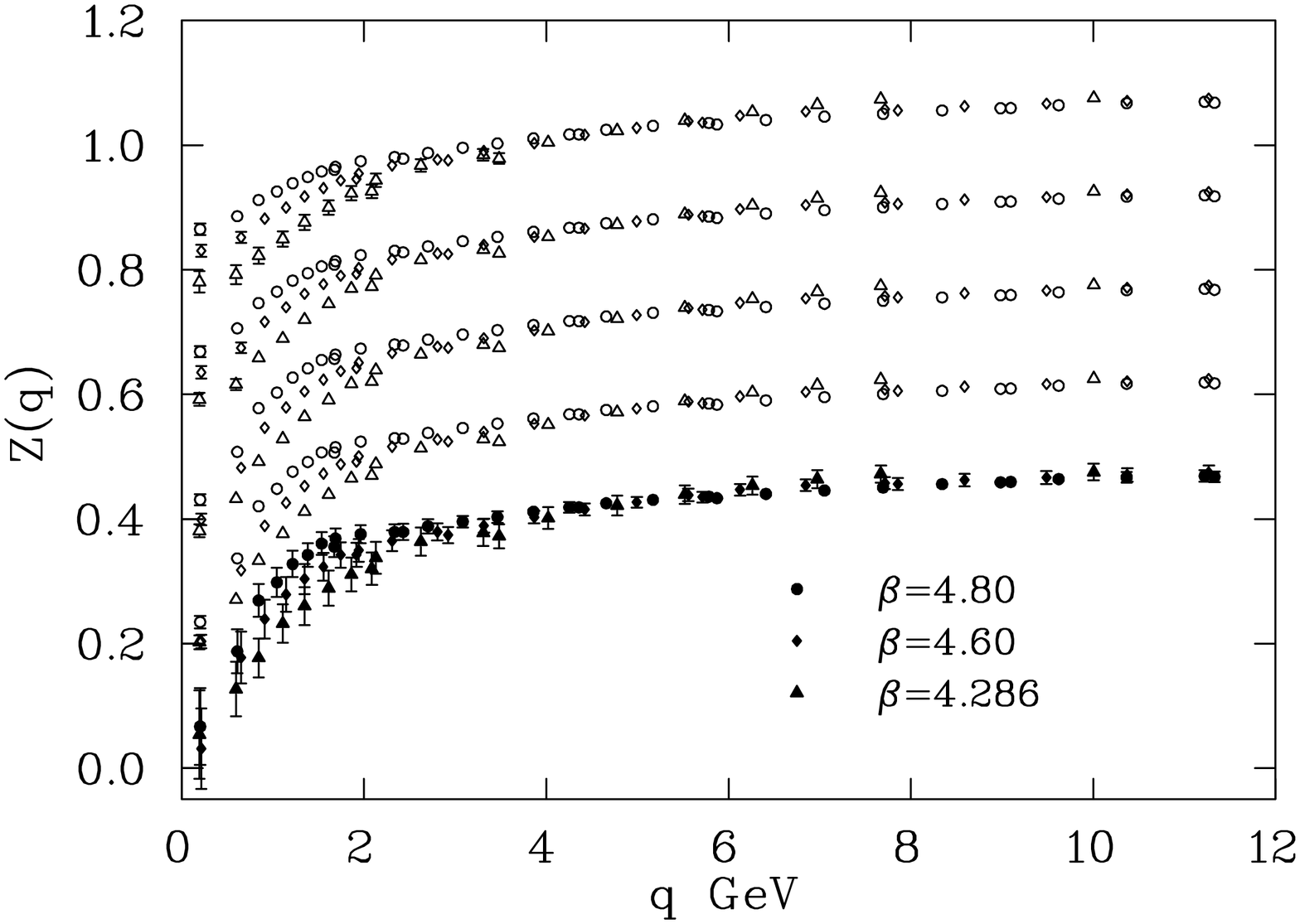}

\includegraphics[angle=90,height=0.28\textheight,width=0.45\textwidth]{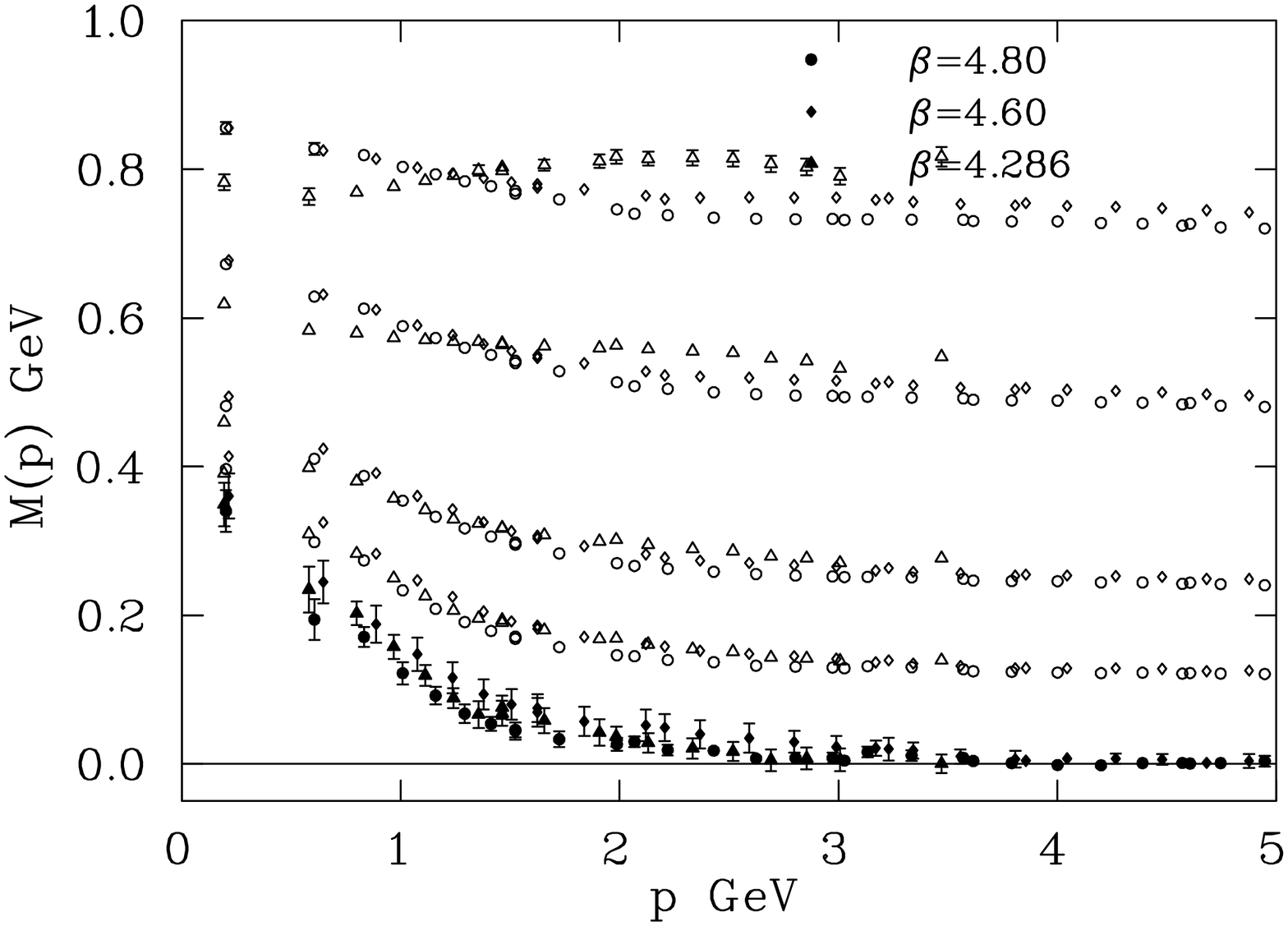}
\includegraphics[angle=90,height=0.28\textheight,width=0.45\textwidth]{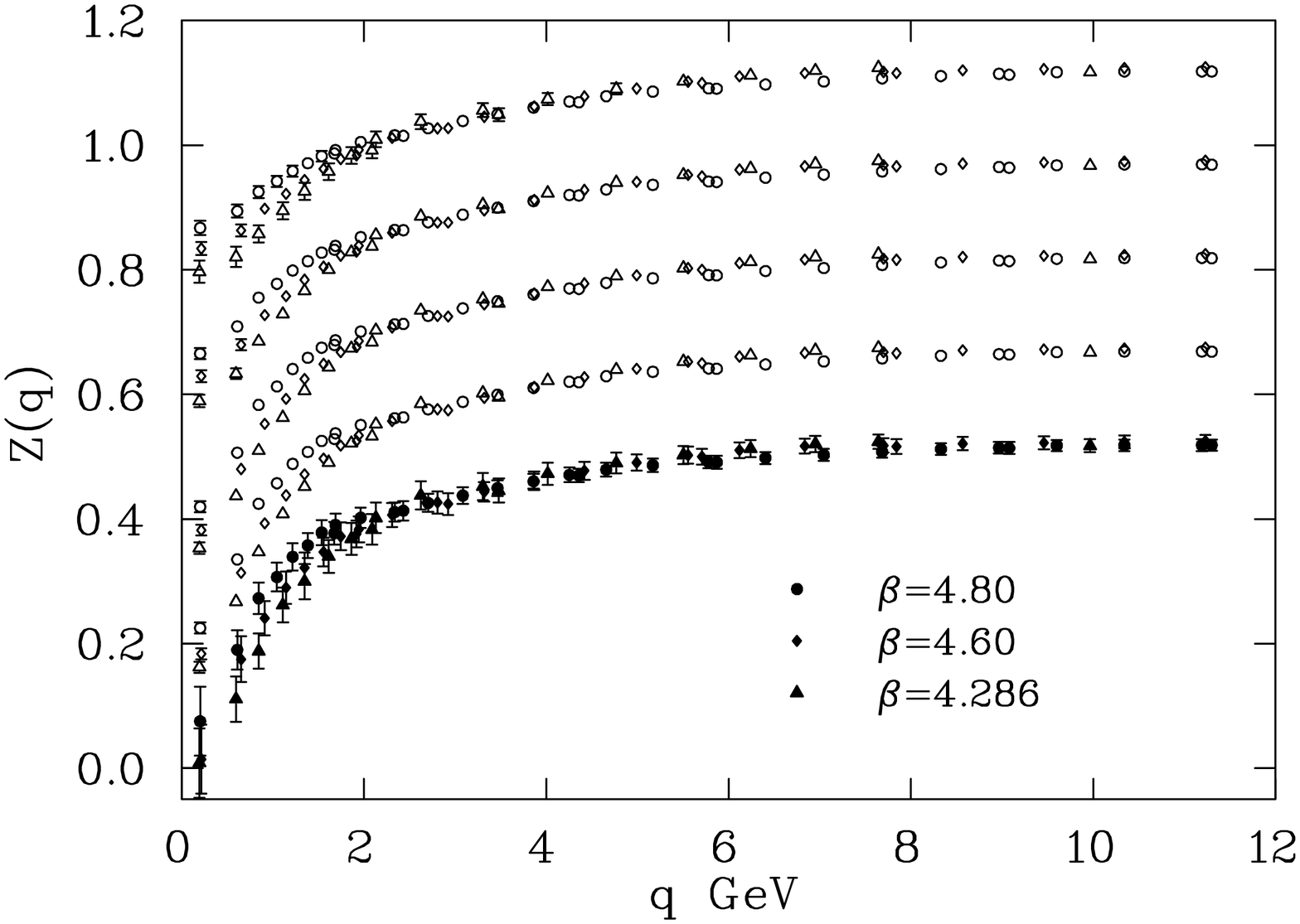}

\caption{Cylinder cut data showing the scaling of the FLIC overlap (top row) and Wilson overlap (bottom row) interpolated mass function $M(p)$ (left) and renormalisation function $Z(p)$ (right) for the three quenched lattices. The different curves from lowest to highest represent matched bare masses of $m^0 = 0.0, 0.1, 0.2, 0.4, 0.6$ GeV. The solid points indicate the chiral limit. The different $Z(q)$ have been offset vertically for clarity.}
\label{fig:contlimit}
\end{figure*}

Using the same technique, we can test the scaling of the mass function and renormalisation function. The results for both the FLIC overlap and Wilson overlap for the three quenched lattices, calculated at $m^0 = 0.0, 0.1, 0.2, 0.4, 0.6$ GeV, are displayed in Fig. \ref{fig:contlimit}. Examining the FLIC Overlap mass function first, we see that the scaling is best at light to medium quark mass. At heavy masses the two finer lattices agree reasonably well, but the coarsest lattice shows scaling violations. At the lighter quark masses the agreement between the lattices is best in the ultraviolet and the very infrared. In the intermediate momentum regime some difference in $M(p)$ between the lattices appears. The difference seems to increase as the quark mass decreases. One possible cause for this difference is in finite volume effects, as the $\beta=4.60$ lattice had a smaller spatial volume than the $\beta=4.80$ lattice. If the smaller volume resulted in an enhancement of the mass function, then this would explain the difference between the finer lattices at lighter masses, as the lighter quark masses would be more sensitive to the small volume. 

The Wilson overlap mass function displays similar scaling properties to the FLIC overlap mass function, but the disagreement between the lattices is in general larger, especially at heavy quark mass. We note that at the lighter quark masses, the FLIC Overlap shows noticably better scaling between $2 < p < 3$ Gev, while the Wilson Overlap has slightly better scaling for $1 < p < 2$ GeV. The Wilson Overlap also appears to scale better in the chiral limit. This could be due to differing finite volume effects. Another, more likely potential source for the difference is in the chiral fits to the data. As mentioned earlier, we use a polynomial fit function. This may be a source of systematic error at light quark mass, as any non-analyticity that might appear in the mass function as we approach the chiral limit would be unaccounted for.

The FLIC overlap renormalisation function $Z(q)$ shows agreement between all three lattices in the ultraviolet, but in the infrared there are differences between the three lattices. Once again, these differences are largest in the intermediate momentum regime. In the very infrared the disagreement is somewhat smaller, in particular at the lighter quark masses. The Wilson overlap renormalisation function again shows similar behaviour, but in the intermediate momentum regime the magnitude of the difference is slightly less. It is somewhat remarkable that for both actions all three lattices agree in the ultraviolet. The tells us that $Z_2(\zeta,a)$ is essentially independent of $a$ (but does differ between the FLIC Overlap and Wilson Overlap).


\section{Conclusion}

We have calculated the quark propagator in momentum space using the FLIC overlap fermion action in quenched QCD. A comparison with a previous study using the Wilson overlap~\cite{overlgp} indicates that both the mass function $M(p)$ and the renormalisation function $Z(q)$ agree well in the infrared regime. The comparison was performed using a fitting technique to match the bare quark masses. The chiral behaviour of $M(p)$ and $Z(q)$ was studied, and clearly demonstrates non-linear behaviour at low momenta and light quark mass, which is consistent with expectations from chiral perturbation theory. 

The heavy quark mass behaviour of the FLIC overlap appears to be better than that of the Wilson overlap, as the FLIC overlap seems to inherit the benefits of smearing the irrelevant operators in the FLIC kernel. As one goes to lighter quark masses, $M(p)$ for the two actions agree even in the ultraviolet. The same is not true of $Z(q),$ as the two actions disagree in the ultraviolet for all masses.

The scaling of the Wilson overlap and FLIC overlap results were also studied. The two actions showed similar scaling behaviour, with some differences. The results for $M(p)$ showed relatively good scaling across the three lattices at $a=0.096, 0.120\text{ and } 0.200.$ The scaling was best in the ultraviolet and in the very infrared, with some differences appearing in the intermediate momentum regime. The results at $a=0.200$ showed clear scaling violations at the heavy quark masses.  Overall, for $M(p)$ we see that the FLIC overlap shows superior scaling behaviour at heavy quark mass and in the ultraviolet. The scaling of the renormalisation function $Z(p)$ was good in the ultraviolet, but disagreements appeared between the three lattices in the infrared. 

Given the superior properties of the FLIC overlap operator outlined above, and the fact that the FLIC overlap operator is significantly cheaper to invert, we advocate the FLIC overlap operator as the operator of choice for chiral studies of hadron phenomenology.
 
Future work will explore the effects of dynamical FLIC fermions~\cite{kamleh-hmc} on the quark propagator~\cite{kamleh-dynflicoverglu}. Further investigations of the potential effects of finite volumes and the chiral extrapolation function used will also be conducted.

\acknowledgments

We thank both the South Australian Partnership for Advanced Computing (SAPAC)
and the Australian Partnership for Advanced Computing (APAC)
for generous grants of supercomputer time which have enabled this
project.  This work is supported by the Australian Research Council.


\end{document}